\documentclass[onecolumn, prd, aps, tightenlines, preprintnumbers, showpacs, nofootinbib, superscriptaddress, notitlepage]{revtex4-1}

\pdfoutput=1

\usepackage{float}
\usepackage{amsmath}
\usepackage{color}
\usepackage{graphicx}
\usepackage[dvipsnames]{xcolor}
\usepackage{url}
\usepackage{epsfig}
\usepackage[T1]{fontenc}
\usepackage{multirow}
\usepackage{physics} % use \bra{...} and \ket{...}, \abs{x}, \norm{x}, \eval{x}_0^\infty, \order{x^2}, comm{A,B}, acomm{A, B}, \vdot, \grad \div \curl, \dd, \pdv{x}, \pdv{f}{x} \pdv{f}{x}{y}, \braket{a}{b}, \expval{A}, \dyad{a}{b}
\usepackage{booktabs} % for much better looking tables
\usepackage{array} % for better arrays (eg matrices) in maths
\usepackage{paralist} % very flexible & customisable lists (eg. enumerate/itemize, etc.)
\usepackage{grffile}
\usepackage{verbatim} % adds environment for commenting out blocks of text & for better verbatim
\usepackage{subfig} % make it possible to include more than one captioned figure/table in a single float
\usepackage{amsmath,amsthm,amssymb,bm,amsfonts}
\usepackage{slashed}
\usepackage[utf8]{inputenc}
\usepackage{hyperref}
\allowdisplaybreaks[1]

\usepackage{color}
\usepackage[normalem]{ulem}
\usepackage{multirow}
\usepackage[title]{appendix}

\def\beg{\begin{equation}}
\def\eeg{\end{equation}}
\def\bea{\begin{eqnarray}}
\def\eea{\end{eqnarray}}

\newcommand{\slv}{\raise.15ex\hbox{$/$}\kern-.53em\hbox{$v$}}
\newcommand{\slnbar}{\raise.15ex\hbox{$/$}\kern-.53em\hbox{$\bar{n}$}}
\newcommand{\slF}{\raise.15ex\hbox{$/$}\kern-.53em\hbox{$F$}}
\newcommand{\sllbar}{\raise.15ex\hbox{$/$}\kern-.40em\hbox{$\bar{l}$}}
\newcommand{\slh}{\raise.15ex\hbox{$/$}\kern-.40em\hbox{$h$}}
\newcommand{\slP}{\raise.15ex\hbox{$/$}\kern-.53em\hbox{$P$}}
\newcommand{\slR}{\raise.15ex\hbox{$/$}\kern-.53em\hbox{$R$}}
\newcommand{\slz}{\raise.15ex\hbox{$/$}\kern-.53em\hbox{$Z$}}
\newcommand{\slzbar}{\raise.15ex\hbox{$/$}\kern-.53em\hbox{$\bar{Z}$}}
\newcommand{\slQ}{\raise.15ex\hbox{$/$}\kern-.53em\hbox{$Q$}}
\newcommand{\slK}{\raise.15ex\hbox{$/$}\kern-.53em\hbox{$K$}}
\newcommand{\slkbar}{\raise.15ex\hbox{$/$}\kern-.53em\hbox{$\bar{k}$}}
\newcommand{\slkone}{\raise.15ex\hbox{$/$}\kern-.53em\hbox{$k_1$}}
\newcommand{\slpone}{\raise.15ex\hbox{$/$}\kern-.53em\hbox{$p_1$}}
\newcommand{\slpbarone}{\raise.15ex\hbox{$/$}\kern-.53em\hbox{$\bar{p}_1$}}
\newcommand{\slptwo}{\raise.15ex\hbox{$/$}\kern-.53em\hbox{$p_2$}}
\newcommand{\slpbartwo}{\raise.15ex\hbox{$/$}\kern-.53em\hbox{$\bar{p}_2$}}
\newcommand{\slqone}{\raise.15ex\hbox{$/$}\kern-.53em\hbox{$q_1$}}
\newcommand{\slD}{\raise.15ex\hbox{$/$}\kern-.53em\hbox{$\!D$}}
\newcommand{\slC}{\raise.15ex\hbox{$/$}\kern-.53em\hbox{$C$}}
\newcommand{\slA}{\raise.15ex\hbox{$/$}\kern-.73em\hbox{$A$}}
\newcommand{\slSigma}{\raise.15ex\hbox{$/$}\kern-.53em\hbox{$\Sigma$}}
\newcommand{\slpartial}{\raise.15ex\hbox{$/$}\kern-.53em\hbox{$\partial$}}
\newcommand{\slcalP}{\raise.15ex\hbox{$/$}\kern-.63em\hbox{$\cal P$}}
\newcommand{\sleps}{\raise.15ex\hbox{$/$}\kern-.53em\hbox{$\epsilon$}}
\newcommand{\slepsbar}{\raise.15ex\hbox{$/$}\kern-.53em\hbox{$\overline{\epsilon}$}}
\newcommand{\slepsstar}{\raise.15ex\hbox{$/$}\kern-.53em\hbox{$\epsilon$}^\star}
\newcommand{\slS}{\raise.15ex\hbox{$/$}\kern-.73em\hbox{$S$}}

\newcommand{\td}{\text{d}}

\newcommand{\bb}{\mathbf}

\newcommand{\be}{\boldsymbol{\epsilon}}
\newcommand{\bk}{\mathbf{k}}
\newcommand{\bp}{\mathbf{p}}
\newcommand{\bq}{\mathbf{q}}
\newcommand{\bx}{\mathbf{x}}
\newcommand{\by}{\mathbf{y}}

\newcommand{\de}{\cdot\boldsymbol{\epsilon}}
\newcommand{\des}{\cdot \boldsymbol{\epsilon}^*}

\newcommand{\dtwo}[1]{\frac{\dd^2 #1}{(2\pi)^2}}

\newcommand{\p}{\prime}

\newcommand{\bz}{\bar{z}}
\newcommand{\eq}[1]{\begin{align} #1 \end{align}}

\newcommand{\pa}[1]{\left( #1 \right)}
\newcommand{\br}[1]{\left[ #1 \right]}

\begin{document}
\title{SIDIS at small $x$ at next-to-leading order: transverse photon}

\author{Tolga Altinoluk}
%\email{...}
\affiliation{Theoretical Physics Division, National Centre for Nuclear Research, Pasteura 7, Warsaw 02-093, Poland}

\author{Filip Bergabo}
%\email{fbergabo@gradcenter.cuny.edu}
\affiliation{Department of  Physics and Mathematics, Jacksonville University,
2800 University Blvd N, Jacksonville, FL 32211, USA}

\author{Jamal Jalilian-Marian}
%\email{jamal.jalilian-marian@baruch.cuny.edu}
\affiliation{
Department of Natural Sciences, Baruch College, CUNY, 17 Lexington Avenue, New York, NY 10010, USA}
\affiliation{City University of New York Graduate Center, 365 Fifth Avenue, New York, NY 10016, USA}

\author{Cyrille Marquet}
%\email{cyrille.marquet@polytechnique.edu}
\affiliation{CPHT, CNRS, Ecole Polytechnique, Institut Polytechnique de Paris, 91120 Palaiseau, France}

\author{Yu Shi}
%\email{yu.shi@polytechnique.edu}
\affiliation{CPHT, CNRS, Ecole Polytechnique, Institut Polytechnique de Paris, 91120 Palaiseau, France}
\affiliation{Key Laboratory of Particle Physics and Particle Irradiation (MOE), Institute of frontier and interdisciplinary science, Shandong University, Qingdao, Shandong 266237, China}

%%%%%%%%%%%%%%%%%%%%%%%%
\begin{abstract}

We calculate the next-to-leading order corrections to single inclusive hadron production in deep inelastic scattering at small $x$ using the color glass condensate formalism, for the case when the exchanged photon is transversely polarized. We show all UV and soft divergences cancel while collinear and rapidity divergences result in scale evolution of quark-hadron fragmentation function and small-x evolution of dipole amplitudes, respectively.

\end{abstract}

\maketitle

\section{Introduction}\label{sec:intro}

The Color Glass Condensate (CGC) (see \cite{Gelis:2010nm, Albacete:2014fwa,Blaizot:2016qgz} for reviews and references therein) is an effective field theory that describes the high-energy, small-$x$ regime of Quantum Chromodynamics (QCD), where gluon densities become very large and nonlinear effects dominate. In this regime, the rapid growth of gluon distributions provided  by linear QCD evolution is expected to be tamed by gluon recombination effects, leading to the phenomenon of gluon saturation. The CGC formalism provides a framework to study these nonlinear dynamics by describing the high-energy hadron in terms of strong classical color fields. This approach has been instrumental in understanding various phenomena observed in high-energy collisions, including deep inelastic scattering (DIS) at Hadron-Electron Ring Accelerator (HERA) and proton-nucleus and heavy-ion collisions collisions at Relativistic Heavy Ion Collider (RHIC) and the Large Hadron Collider (LHC). By incorporating nonlinear evolution equations, such as the Balitsky-Kovchegov (BK) \cite{Balitsky:1995ub, Kovchegov:1999yj, Kovchegov:1999ua} and Jalilian-Marian–Iancu–McLerran–Weigert–Leonidov–Kovner (JIMWLK) \cite{Jalilian-Marian:1996mkd,Jalilian-Marian:1997qno,Jalilian-Marian:1997jhx,Jalilian-Marian:1997ubg,Kovner:2000pt,Weigert:2000gi,Iancu:2000hn,Iancu:2001ad,Ferreiro:2001qy} equations, the CGC framework offers a systematic way to explore the transition from a dilute to a saturated system providing crucial insights into the initial conditions of heavy-ion collisions. 

The search for gluon saturation and the verification of its role in high-energy QCD dynamics are among the central objectives of current and future collider experiments. Presently, measurements at RHIC and the LHC have provided tantalizing hints of saturation effects in proton-nucleus and nucleus-nucleus collisions, such as the suppression of forward hadron production and the modification of dihadron correlations. However, definitive confirmation of saturation physics requires more precise experimental data and theoretical advancements. Future high-energy facilities, such as the Electron-Ion Collider (EIC) in the United States and the proposed Large Hadron Electron Collider (LHeC) at CERN, will offer unprecedented opportunities to explore the small-$x$ regime with DIS on heavy nuclei, providing a cleaner and more controlled environment to probe saturation phenomena. In order to fully utilize these experimental programs, the precision of the theoretical calculations have to be increased. 

Within the CGC framework, DIS related observables are usually studied using the dipole factorization \cite{Bjorken:1970ah,Nikolaev:1990ja}. In this factorization framework, the incoming virtual photon splits into a quark-antiquark pair which then scatters off the dense target. In order to improve the theoretical precision, next-to-leading order (NLO) corrections in strong coupling constant $\alpha_s$ to DIS related observables have been studied for almost a decade now. Inclusive DIS has been computed at NLO with massless quarks in \cite{Balitsky:2010ze,Balitsky:2012bs,Beuf:2011xd,Beuf:2016wdz,Beuf:2017bpd,Ducloue:2017ftk,Hanninen:2017ddy} and its fits to HERA data is performed in \cite{Beuf:2020dxl}. The same study is also performed for massive quarks in \cite{Beuf:2021qqa,Beuf:2021srj,Beuf:2022ndu} with its fit to HERA data provided in \cite{Hanninen:2022gje}. Structure functions in diffractive DIS have been computed at NLO for massless quarks in \cite{Beuf:2022kyp,Beuf:2024msh}. Apart from the studies that focus on the NLO corrections to the structure functions, inclusive dijet \cite{Caucal:2021ent,Taels:2022tza,Caucal:2022ulg,Caucal:2023nci,Caucal:2023fsf} and dihadron \cite{Ayala:2016lhd,Ayala:2017rmh,Bergabo:2023wed,Bergabo:2022tcu,Iancu:2022gpw,Caucal:2024nsb} as well as diffractive dijet \cite{Boussarie:2016ogo,Boussarie:2019ero} and dihadron \cite{Fucilla:2022wcg} production in DIS have been computed at NLO. The computations performed to study the production cross section of dihadrons/dijets at NLO allows one to obtain the NLO cross section for single inclusive hadron and jet production by integrating over one of the produced hadron/jet. Inclusive and diffractive single hadron and single jet production have been studied both in general kinematics \cite{Fucilla:2023mkl,Caucal:2024cdq, Bergabo:2024ivx, Bergabo:2022zhe} and in the large virtuality limit \cite{Altinoluk:2024vgg, Caucal:2024vbv, Caucal:2025qjg} at NLO. Apart from the aforementioned studies on inclusive and diffractive jet/hadron production at NLO, various different aspects of single and multi jet/hadron production have been studied within the CGC framework \cite{Kovchegov:2001ni, Marquet:2004xa, Golec-Biernat:2005prq, Marquet:2009ca,Altinoluk:2015dpi, Hatta:2016dxp, Mantysaari:2019csc, Salazar:2019ncp, Iancu:2020jch, Iancu:2021rup,Hatta:2022lzj, Iancu:2022lcw, Iancu:2023lel, Rodriguez-Aguilar:2023ihz, Kar:2023jkn, Hauksson:2024bvv, Taels:2023czt}.  

In this manuscript, we study the single inclusive hadron production in DIS via transverse photon exchange at NLO starting from the NLO dihadron production cross section. We integrate out the antiquark and let the produced quark fragment into a hadron. The manuscript is organized as follows. In Section \ref{LO_X_section}, we briefly summarize the leading order cross section both for double and single inclusive production in DIS at partonic level. In Section \ref{NLO_X_section}, we compute the NLO corrections to the single inclusive quark production cross section in DIS via transverse photon exchange. Section \ref{divergences} is devoted to the discussion of the cancellation of the divergences that appear at NLO and evolution of the fragmentation functions. Finally, in Section \ref{conc}, we summarize our results and provide an outlook. In Appendix \ref{app:dihadron_NLO}, we provide the quark-antiquark production cross section in DIS via transverse photon at NLO and at finite $N_c$. In Appendix \ref{App:z_to_zero}, the $z\to 0$ limit of the single inclusive quark production in DIS for the transversely polarized photon is presented. These results are provided at finite $N_c$ and used to extract the rapidity divergences.

\section{Leading-order partonic cross sections}
\label{LO_X_section}

 In the small $x$ limit of DIS and in the target rest frame the virtual photon (transverse or longitudinal) splits into a quark-antiquark pair, which then multiply scatters off the target hadron or nucleus. At leading order, the partonic production cross section of a quark with transverse momentum $\bp$, rapidity $y_1$ and an antiquark with transverse momentum $\bq$ and rapidity $y_2$ reads  
\bea
\frac{\dd \sigma^{\gamma^*A \to q\bar{q} X}}
{\dd^2 \bb{p}\, \dd^2 \bb{q} \, \dd y_1 \, \dd y_2} &=& 
\frac{ e^2 Q^2 N_c}{(2\pi)^7} \delta(1-z_1-z_2) \, (z_1 z_2)^2 \, 
\int \dd^8 \bx \left[S_{122^\prime 1^\prime} - S_{12} - S_{2^\prime 1^\prime} + 1\right]\,  
e^{i\bp\cdot\bx_{1^\p1}} e^{i\bq\cdot\bx_{2^\p2}} 
\nonumber \\
&  
\times & \, \bigg\{4z_1z_2K_0(|\bb{x}_{12}|Q_1)K_0(|\bb{x}_{1^\prime 2^\prime}|Q_1) +  
(z_1^2 + z_2^2) \,
\frac{ \bb{x}_{12}\cdot \bb{x}_{1^\prime 2^\prime}}{|\bb{x}_{12}| |\bb{x}_{1^\prime 2^\prime}|} \, 
K_1(|\bb{x}_{12}|Q_1)K_1(|\bb{x}_{1^\prime 2^\prime}|Q_1) 
\bigg\} .\label{LOdsig}
\eea
where the first and second terms in the curly bracket above correspond to the contribution of the longitudinal and transverse polarizations of the incoming photon with virtuality $Q^2$. Here, $K_\alpha(\cdots)$ is the modified Bessel function of the second type. We have also defined  
\begin{align}
Q_i = Q\sqrt{z_i(1-z_i)}, \,\,\,\,\,\, \bx_{ij} = \bx_i - \bx_j,\,\,\,\,\,\, \dd^8 \bx = \dd^2 \bx_1 \, \dd^2 \bx_2\, \dd^2 \bx_{1^\p} \, \dd^2 \bx_{2^\p}.
\end{align}
The production cross section is a convolution of the probability for a photon to split into a quark at transverse position $\bx_1$ and an antiquark at position $\bx_2$ represented by the Bessel functions, with the probability for this quark antiquark pair to multiply scatter from the target which is encoded in the dipoles $S_{ij}$ and quadrupoles $S_{ijkl}$. Dipoles and quadrupoles are two-point and four-point correlators of Wilson lines defined as 
\begin{eqnarray}
    S_{i j} &\equiv& S (\bx_i, \by_j) = \frac{1}{N_c} 
    \big\langle \text{tr} \big[ V (\bx_i) V^\dagger (\by_j)\big] \big\rangle
    \nonumber \\
    S_{i j k l} &\equiv& S (\bx_i, \bx_j, \bx_k, \bx_l) = 
    \frac{1}{N_c}
    \big\langle \text{tr}\big[ V (\bx_i) V^\dagger (\bx_j) V (\bx_k) V^\dagger (\bx_l)\big] \big\rangle 
\end{eqnarray}
where the Wilson line $V ( \bx_i)$ is a path ordered exponential of the gluon field in the fundamental representation. It efficiently resums multiple soft scatterings of the quark projectile on the target proton or nucleus which is modeled as a  classical color field. We note that this description corresponds to making the well-known eikonal approximation which is not unique to CGC formalism but is also used in calculations of energy loss effects from a dense medium (see e.g. \cite{Kovner:2003zj,Munier:2016oih}). Furthermore, the virtual photon has momentum $l^\mu$ with $l^2 = -Q^2$ and we have set the transverse momentum of the photon to zero without any loss of generality. Also, $p^\mu$ ($q^\mu$) is the momentum of the outgoing quark (antiquark) and $z_1$ ($z_2$) is its longitudinal momentum fraction relative to the photon. Transverse coordinate of quark (antiquark) is denoted $\bx_1$ ($\bx_2$) and primed coordinates are used in the conjugate amplitude. Quark and antiquark rapidities $y_1$ and $y_2$ are related to their momentum fractions $z_1$ and $z_2$ via $\dd y_i = \dd z_i / z_i$. To calculate the single inclusive production cross section, we integrate over one of the final state partons phase space; we choose it to be the antiquark so that the quark is produced. This sets $\bx_2 = \bx_{2^\p}$ and gives 
\bea
\frac{\dd \sigma_T^{\gamma^*A \to q (\bp,y_1)X}}{\dd^2 \bp\, \dd y_1} =
\frac{ e^2 Q^2 N_c}{(2\pi)^5} 
z_1^2 \bz_1 (z_1^2 + \bz_1^2) 
\int \dd^6 \bx \left[S_{11^\p} - S_{12} - S_{1^\p2} + 1\right] 
e^{i\bp\cdot\bx_{1^\p1}}
\frac{\bx_{12}\cdot \bx_{1^\p2}} {|\bx_{12}||\bx_{1^\p2|}} \, 
K_1(|\bx_{12}|Q_1) K_1(|\bx_{1^\p2}|Q_1)
\label{LOdsig-sidis-quark}
\eea
where we only kept the transverse-photon part and with $\bz_1=1-z_1$.
The only non-trivial ingredients of the cross section above are the dipoles $S_{ij}$ that contain the QCD dynamics at small $x$. While the LO cross section can provide us with a qualitative and even a semi-quantitative QCD dynamics in a proton or nucleus, it is imperative to go beyond a semi-quantitative analysis and improve the precision of the theoretical calculations in preparation for the upcoming Electron-Ion Collider (EIC). There is already a significant amount of work done towards calculating next-to-leading order (NLO) corrections to the BK-JIMWLK evolution equations which is a major step in this direction. Here we calculate the NLO corrections to the single inclusive hadron production in DIS (SIDIS) for the case when the exchanged photon is transversely polarized. 
%~\footnote{Next to leading order corrections to SIDIS for longitudinal photon exchange is calculated in~\cite{Bergabo:2022zhe}}.

\section{Next to leading order corrections}
\label{NLO_X_section}

\begin{figure}[H]
\centering
\includegraphics[width=70mm]{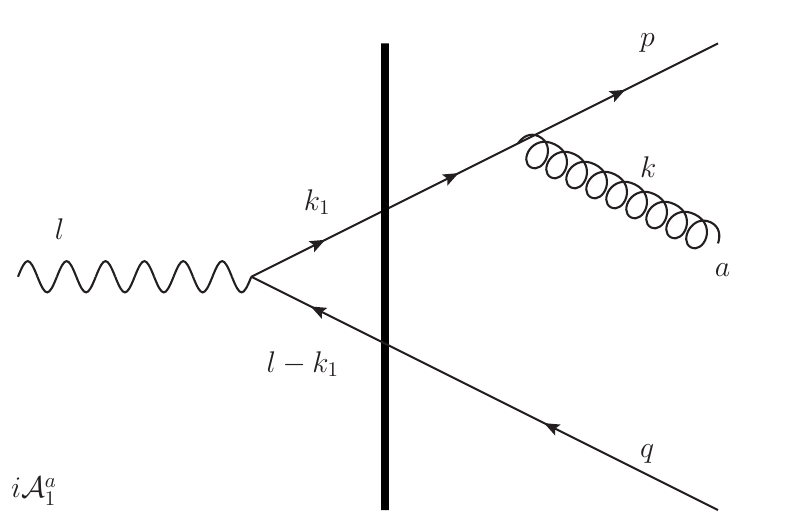}\includegraphics[width=70mm]{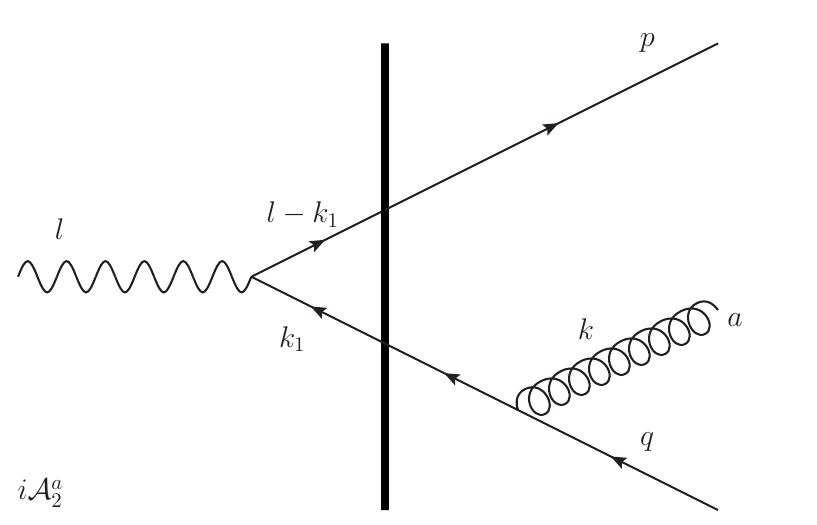}\\ \includegraphics[width=70mm]{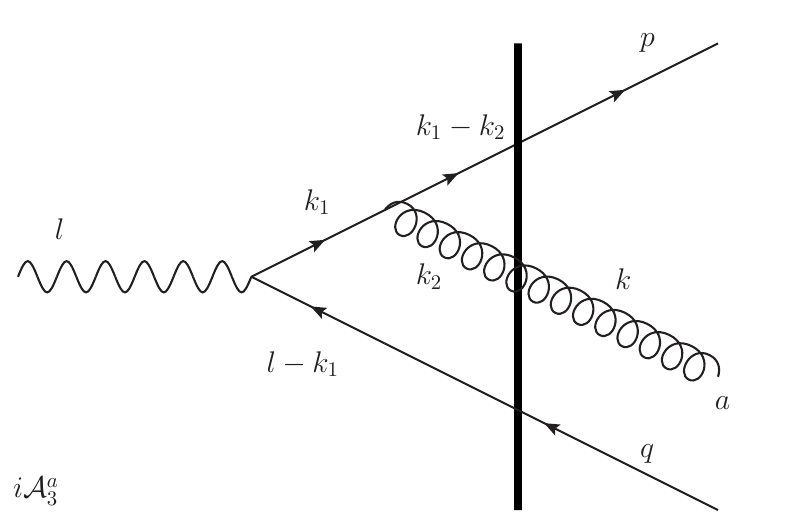}\includegraphics[width=70mm]{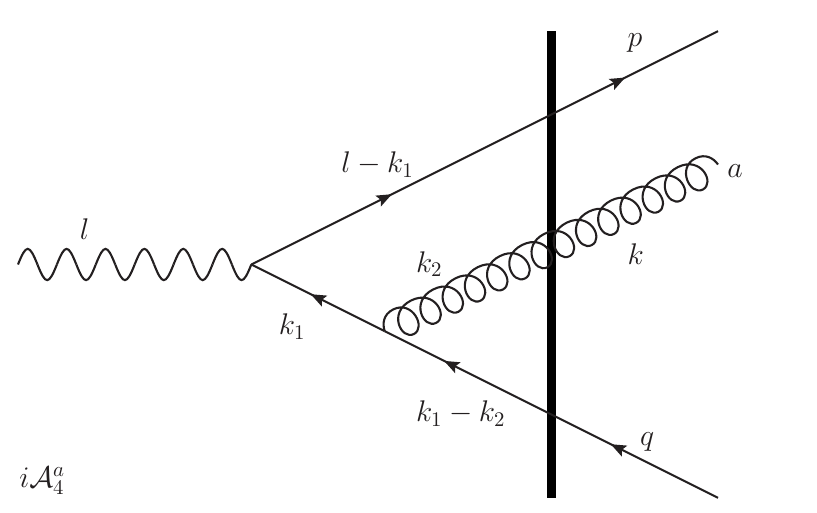}
\caption{The real corrections $i\mathcal{A}_1^a, ..., i\mathcal{A}_4^a$. The arrows on Fermion lines indicate Fermion number flow, all momenta flow to the right. The thick solid line indicates interaction with the target.}\label{fig:realdiags}
\end{figure}

The complete expressions for NLO corrections to double inclusive quark-antiquark production with transversely polarized exchanged photon were derived in~\cite{Bergabo:2023wed} and written down in the large-$N_c$ limit. Here, we start from those expressions but including the full $N_c$ structure; they are given in Appendix A.
In order to calculate the NLO corrections to SIDIS we start with those and integrate out the antiquark phase space, as was done at LO. There are both real and virtual corrections in the amplitude; real corrections involve radiation of a gluon either before or after scattering from the target where the gluon can be radiated from either quark or antiquark in the amplitude as shown in Fig.~\ref{fig:realdiags}, with $z$ and $\bk$ denoting the gluon longitudinal momentum fraction and transverse momentum, respectively. The cross section is then obtained from multiplying the amplitude with the complex conjugate amplitude. 

\begin{figure}[H]
\centering
\includegraphics[width=60mm]{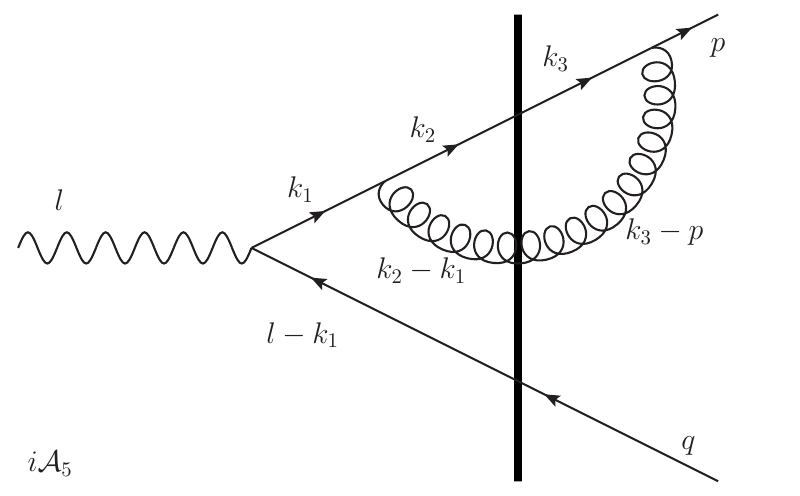}\includegraphics[width=60mm]{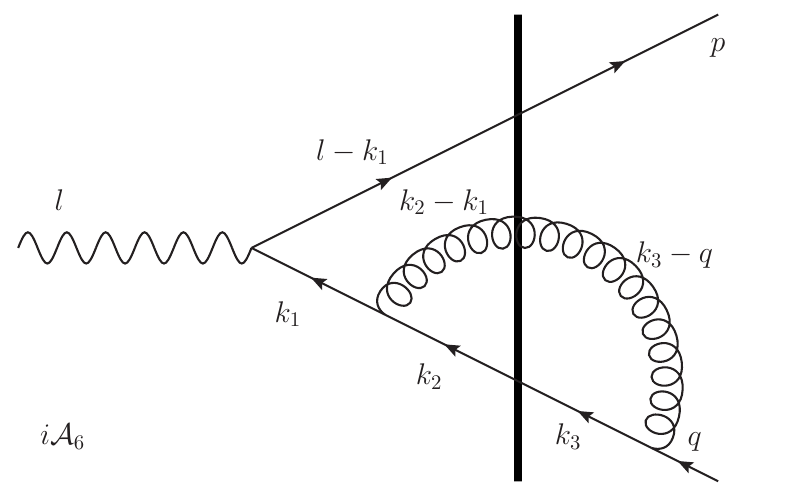}\\ \includegraphics[width=60mm]{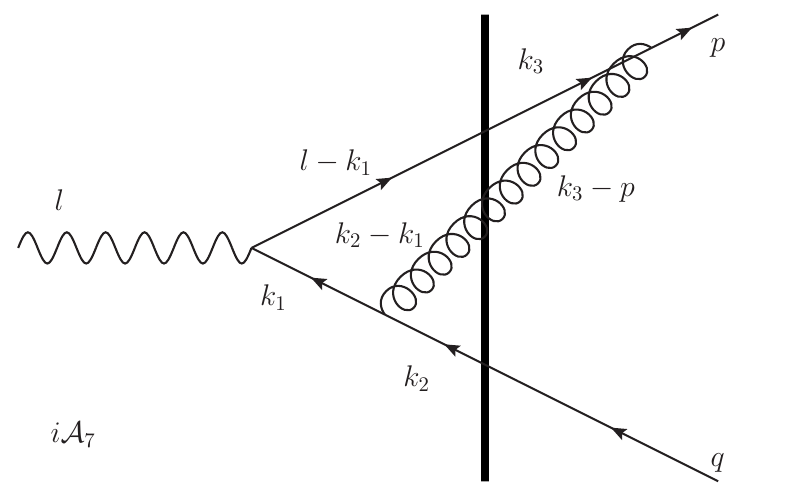}\includegraphics[width=60mm]{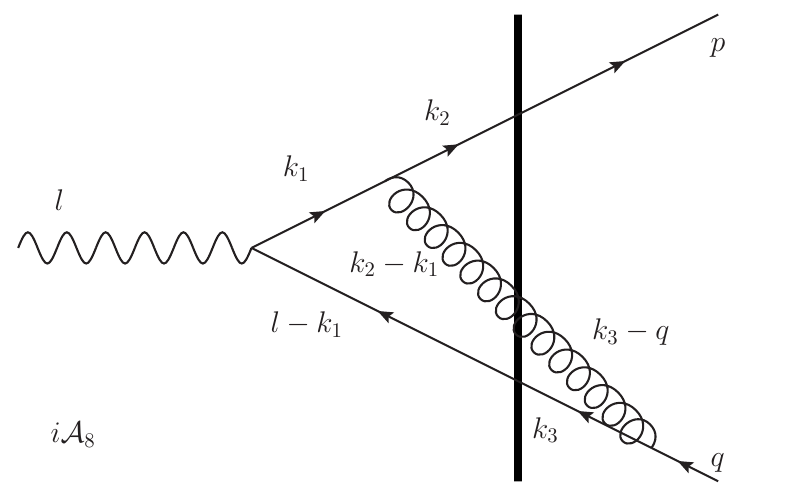}\\ \includegraphics[width=60mm]{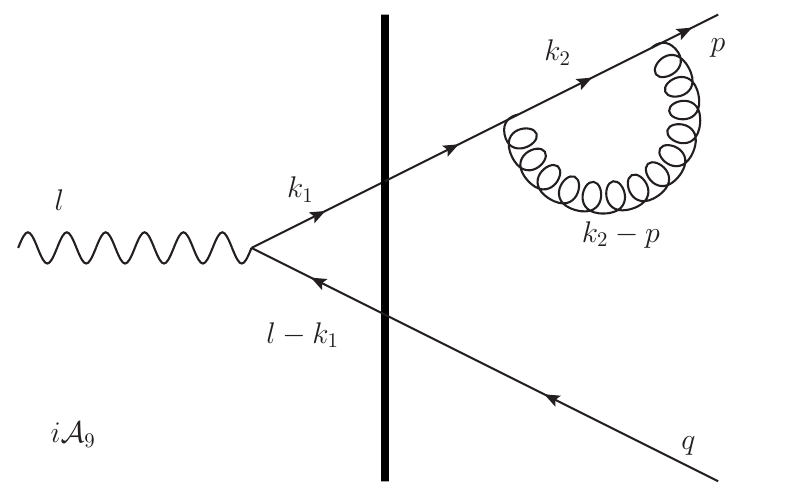}\includegraphics[width=60mm]{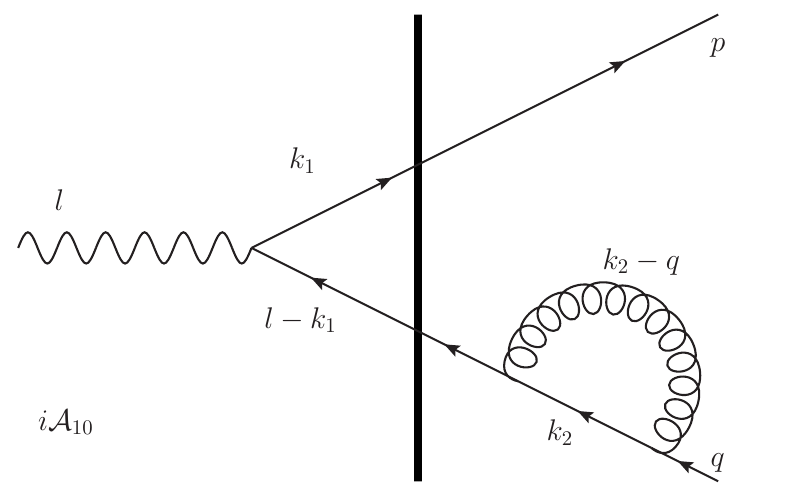}\\ \includegraphics[width=60mm]{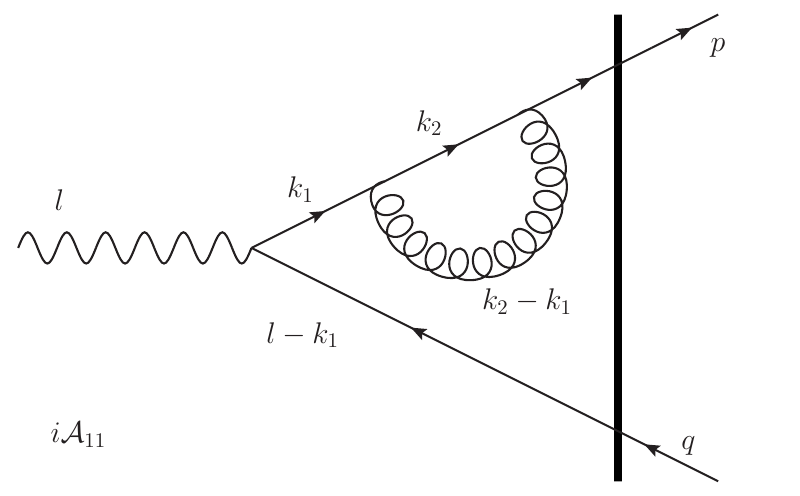}\includegraphics[width=60mm]{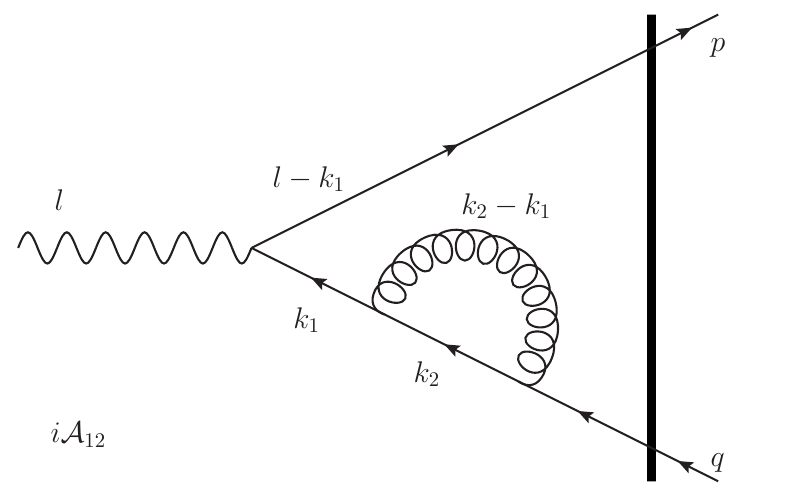}\\ \includegraphics[width=60mm]{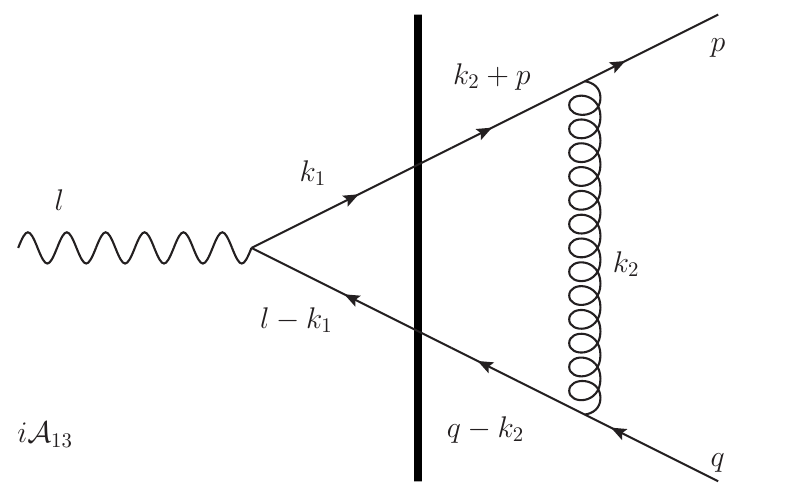}\includegraphics[width=60mm]{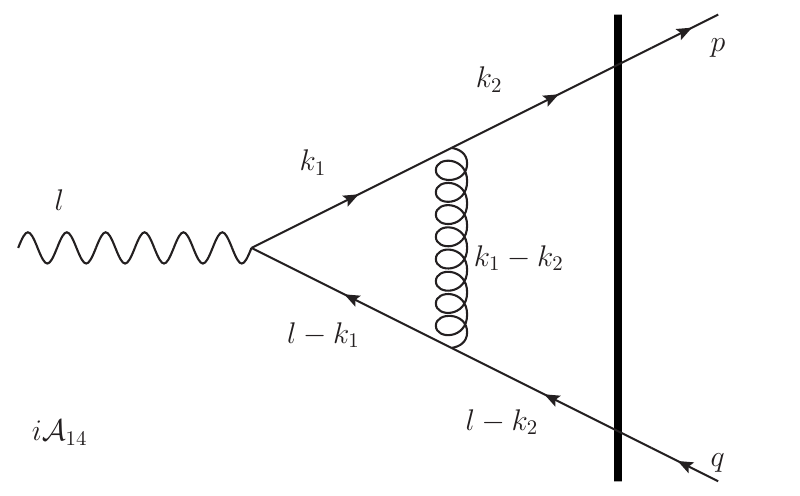}
\caption{The ten virtual NLO diagrams $i\mathcal{A}_5, ..., i\mathcal{A}_{14}$. The arrows on fermion lines indicate fermion number flow, all momenta flow to the right, \textit{except} for gluon momenta. The thick solid line indicates interaction with the target.}\label{virtualdiags}
\end{figure}

The virtual corrections to the amplitude also involve radiation of a gluon from quark or antiquark and either before or after scattering from the target. However in this case the radiated gluon is then absorbed by either the quark or antiquark, before or after scattering from the target at the amplitude (or complex conjugate amplitude) level already. One then obtains the corresponding cross section by multiplying this amplitude with the LO complex conjugate amplitude (or vice versa). The diagrams for virtual corrections to the amplitude are shown in Fig.~\ref{virtualdiags}, where the gluon momentum labeling is now diagram dependent. In our expressions for the virtual diagrams, the transverse part of the gluon momentum will follow those notations but we shall keep using $z$ to denote the gluon longitudinal momentum fraction.  
%%%%%%
To summarize, by following the same conventions introduced in \cite{Bergabo:2022tcu}, we write the total NLO contribution to the single inclusive quark production in DIS via transverse photon as 
\begin{align}
\label{eq:NLO_X_sec_Sym}
\td\sigma^T_{NLO}=\sum_{i=1}^4\td\sigma_{i\times i}+\sum_{i=2}^4\sum_{j=1}^{i-1} \Big(\td\sigma_{i\times j}+\td\sigma_{j\times i}\Big)+\sum_{i=5}^{14}\Big(\td\sigma_i+\td\sigma^\star_i\Big)
\end{align}
where the first two terms are the real contributions which can be computed as  
\begin{align}
\td\sigma_{i\times j}=\frac{1}{\cal F}\int_{z,\bk,z_2, \bq}
%\td z\ \frac{\td^2 \bk}{(2\pi)^2} \, \td z_2 \, \frac{\td^2\bq}{(2\pi)^2}
\Big[(i{\cal M}^a_i)(i{\cal M}^a_j)^\star\Big]\td \Phi^{(3)}
\end{align}
Here, ${\cal M}_i$ corresponds to the real amplitudes computed in \cite{Bergabo:2023wed} by using the helicity amplitude method and shown in Fig. \ref{fig:realdiags}. ${\cal F}$ is the incident flux which is taken to be $2l^+$ and $\td\Phi^{(3)}$ corresponds to the phase space factor and reads \cite{Bergabo:2022tcu}
\begin{align}
\td \Phi^{(3)} = 2l^+ \frac{\td^2\bp\, \td^2 \bq\, \td^2 \bk\, \td y_1 \,  \td y_2\, \td z}{(2\pi)^8 (4l^+)^2z} \, \delta(1-z_1-z_2-z) 
\end{align}
On the other hand, the last term in Eq. \eqref{eq:NLO_X_sec_Sym} correspond to the virtual contributions and those read 
\begin{align}
 \td \sigma_i=\frac{1}{\cal F}\int_{z_2,\bq}\Big[(i{\cal M}_i)(i{\cal M}_0)^\star\Big] \td \Phi^{(2)}
\end{align}% 
where ${\cal M}_0$ is the leading order quark-antiquark amplitude. Here, ${\cal M}_i$ correspond to virtual amplitudes shown in Fig. \ref{virtualdiags} and the phase factor $\td\Phi^{(2)}$ is given by 
\begin{align}
\td\Phi^{(2)}=2l^+\, \frac{\td^2\bp \, \td^2\bq\,\td y_1 \,  \td y_2}{2(2\pi)^5(2l^+)^2}\delta(1-z_1-z_2)
\end{align}

\subsection{Single inclusive hadron (quark) production}

As mentioned above, to calculate single inclusive hadron production in DIS we start with the expressions for double inclusive quark-antiquark production and integrate out the antiquark momenta ($\bq , z_2$) 
so that the quark is produced~\footnote{Note that the result is the same whether one integrates out quark or antiquark, up to a trivial relabeling of momenta.} and hadronizes. After integrating out the antiquark phase space one can see that there are cancellations among some of the diagrams which become topologically equivalent. Specifically the following  diagrams totally cancel after integrating over the antiquark momenta 
\begin{align}
\dd\sigma_{2\times 2} + 2\ \mbox{Re}(\dd\sigma_{10}) = 0 \\
\dd\sigma_{2\times 3} +  \dd\sigma_8^{*} = \dd\sigma_{3\times 2} +  \dd\sigma_8 = 0 \\
\dd\sigma_{2\times 4} +  \dd\sigma_6^{*}  =  
\dd\sigma_{4\times 2} +  \dd\sigma_6  = 0
\label{fullcancel}
\end{align}
This cancellation is easy to understand; after integrating out the phase space of antiquark in quark antiquark production cross section some of the real and virtual diagrams become topologically equivalent except for a relative minus sign that leads to the cancellations. The origin of this relative minus sign is due to the vertex factors $i g$ in the amplitude and and $- i g$ in the complex conjugate amplitude in the NLO real diagrams vs $(i g)^2$  in the amplitude or $(- i g)^2$ in the complex conjugate amplitude in NLO virtual diagrams.

\subsubsection{Real corrections to single inclusive quark production at NLO}

In this subsection, we provide the the real NLO corrections to the single inclusive quark production that are obtained by starting from the real NLO corrections to the quark-antiquark production given in \cite{Bergabo:2023wed} and integrating over the antiquark momenta $(\bq,z_2)$.  
\begin{align}
\label{def:1x1}
\frac{\text{d} \sigma_{1\times 1}^T}{\text{d}^2 \bp \, \text{d} y_1} &= 
\frac{e^2 g^2 Q^2 N_c}{(2\pi)^{8}}\, 
\int^{\bz_1}_0 \frac{\text{d} z}{z} 
\frac{(\bz_1-z) (z_1 + z)}{z_1}  
\big[ z_1^2+(z_1+z)^2\big]
\big[(\bz_1-z)^2+(z_1+z)^2\big]
%\left[z_1^2 (\bz_1-z)^2 + \left[z_1^2 + (\bz_1-z)^2\right] (z_1+ %z)^2 + (z_1 + z)^4\right] 
\nonumber \\
& \times
\int \text{d}^{8}\bx \, C_F [S_{11^\p} - S_{12} - S_{21^\p} +1] 
K_1(|\bx_{12}|Q_{1z})K_1(|\bx_{1^\p 2}|Q_{1z}) 
\frac{\bx_{12}\cdot\bx_{1^\p 2}}{|\bx_{12}||\bx_{1^\p 2}|}
e^{i\frac{z_1 + z}{z_1}\bp\cdot\bx_{1^\p 1}}\Delta^{(3)}_{1^\p 1}. 
\end{align}
\begin{align}
\label{def:1x2}
&\frac{\td \sigma_{1\times 2}^T}{\td^2 \bp \, \td y_1} =
\frac{e^2 g^2 Q^2 N_c}{(2\pi)^{8}} 
\int^{\bz_1}_0\frac{\td z}{z}
\frac{(\bz_1-z)^2}{\bz_1^2}
%\frac{z_2^2}{(z_2 + z)^2} 
\frac{\sqrt{z_1\bz_1(\bz_1-z)(z_1+z)}}{(\bz_1-z)}
%\frac{\sqrt{z_1 z_2 (1-z_1)(1-z_2)}}{z_2}
\int\td^{8} \bx \; K_1(|\bx_{12}|Q_{1z}) \; K_1(|\bx_{1^\p 2^\p}|Q_1) \nonumber\\
& \times
\bigg[C_F \Big(S_{12} S_{2^\p1^\p} - S_{12} - S_{2^\p1^\p} + 1\Big) 
- \frac{1}{2 N_c} \Big(S_{122^\p1^\p} - S_{12} S_{2^\p1^\p}\Big)\bigg] 
4\Re \bigg\{ \frac{(\bx_{12}\de)(\bx_{1^\p 2^\p}\des)}{|\bx_{12}||\bx_{1^\p2^\p}|}
\\
& \times
\bigg[ \Big(z_1^2+(\bz_1-z)^2\Big)\bz_1(z_1+z)
%\Bigg\{ (z_1^2+z_2^2)(1-z_1)(1-z_2)
\frac{(\bx_{31}\de)(\bx_{2^\p 3}\des)}{\bx_{31}^2 \bx_{2^\p 3}^2} %\nonumber \\
%&
+
\Big(\bz_1^2+(z_1+z)^2\Big)z_1(\bz_1-z)
\frac{(\bx_{31}\des)(\bx_{2^\p 3}\de)}{\bx_{31}^2 \bx_{2^\p 3}^2}\bigg]\bigg\}
e^{i\bp\cdot (\bx_{1^\p 1} + \frac{z}{z_1} \bx_{31})}.\nonumber
\end{align}
\begin{align}
\label{def:3x3}
&\frac{\td \sigma_{3\times 3}^{T}}{\td^2 \bp \, \td y_1}  =
\frac{e^2 g^2 Q^2 N_c}{(2\pi)^{8}}
\int^{\bz_1}_0\frac{\td z}{z} z_1 (\bz_1-z)^2 \, 
\int \td^{8}\bx  \left[\frac{K_1(QX)K_1(QX^\p)}{XX^\p}\right]_{\bx_2^\p = \bx_2} \,
e^{i\bp\cdot\bx_{1^\p 1}} \nonumber \\
& \times 
\Big[
C_F \Big(S_{11^\p} - S_{13} S_{32} - S_{31^\p} S_{23} + 1\Big) 
- \frac{1}{2 N_c} \Big(S_{13} S_{32} + S_{31^\p} S_{23} - S_{12} - S_{21^\p}\Big)
\Big]
\nonumber \\
& \times  
4\Re \bigg\{\bigg[\Big(z_1^2+(\bz_1-z)^2\Big)
\frac{(\bx_{31}\de)(\bx_{3 1^\p}\des)}{\bx_{31}^2\bx_{3 1^\p}^2} 
+\Big((z_1+z)^2 + \frac{z_1^2(\bz_1-z)^2}{(z_1+z)^2}\Big) 
\frac{(\bx_{31}\des)(\bx_{3 1^\p}\de)}{\bx_{31}^2\bx_{3 1^\p}^2}
\bigg]
\nonumber\\
&
\hspace{1.5cm}\times
\big[(z_1\bx_{12}+z\bx_{32})\de\big]
\big[(z_1\bx_{1^\p 2}+z\bx_{3 2})\des\big] \nonumber \\
% &+\Big[(z_1+z)^2 + \frac{z_1^2(\bz_1-z)^2}{(z_1+z)^2}\Big] 
% \frac{(\bx_{31}\des)(\bx_{3 1^\p}\de)}{\bx_{31}^2\bx_{3 1^\p}^2}
% \big[(z_1\bx_{12}+z\bx_{32})\de\big]
% \big[(z_1\bx_{1^\p 2}+z\bx_{3 2})\des\big] \nonumber \\
&
\hspace{1.4cm}
-\frac{z_1^2 z (\bz_1-z)}{2(z_1+z)^2}
\bigg[ \frac{(\bx_{31}\des)}{\bx_{31}^2}\big[(z_1\bx_{12}+z\bx_{32})\de\big] + 
\frac{(\bx_{31^\p}\de)}{\bx_{31^\p}^2}\big[(z_1\bx_{1^\p2}+z\bx_{32})\des\big]\bigg] +
\frac{z_1^2 z^2}{4(z_1+z)^2}\bigg\}.
\end{align}
\begin{align}
\label{def:4x4}
&\frac{\td \sigma_{4\times 4}^{T}}{\td^2 \bp\, \td y_1} =
\frac{e^2 g^2 Q^2 N_c}{(2\pi)^{8}}
\int^{\bz_1}_0\frac{\td z}{z} z_1^3 
\int \td^{8}\bx  \left[\frac{K_1(QX)K_1(QX^\p)}{XX^\p}\right]_{\bx_2^\p = \bx_2} 
e^{i\bp\cdot\bx_{1^\p 1}} \nonumber \\
& \times\Big[
C_F \Big(S_{11^\p} - S_{13} S_{32} - S_{31^\p} S_{23} + 1\Big) 
- \frac{1}{2 N_c} \Big(S_{13} S_{32} + S_{31^\p} S_{23} - S_{12} - S_{21^\p}\Big)\Big]
\nonumber \\
&  \times
4\Re \bigg\{
\bigg[\Big(z_1^2+(\bz_1-z)^2\Big)\frac{(\bx_{32}\de)(\bx_{3 2}\des)}{\bx_{32}^2\bx_{3 2}^2} +\Big(\bz_1^2+\frac{z_1^2(\bz_1-z)^2}{\bz_1^2}\Big)\frac{(\bx_{32}\des)(\bx_{3 2}\de)}{\bx_{32}^2\bx_{3 2}^2}\bigg]
\nonumber\\
&
\hspace{1.2cm}\times
\Big[\Big((\bz_1-z)\bx_{21}+z\bx_{31}\Big)\de\Big]
\Big[\Big((\bz_1-z)\bx_{2 1^\p}+z\bx_{3 1^\p}\Big)\des\Big] \nonumber \\
&-\frac{(\bz_1-z)^2 z_1 z}{2\bz_1^2}
\bigg[ \frac{(\bx_{32}\des)}{\bx_{32}^2}\Big[\Big((\bz_1-z)\bx_{21}+z\bx_{31}\Big)\de\Big] + 
\frac{(\bx_{32}\de)}{\bx_{32}^2}\Big[\Big((\bz_1-z)\bx_{21^\p}+z\bx_{31^\p}\Big)\des\Big]\bigg] 
+\frac{(\bz_1-z)^2 z^2}{4(z_1+z)^2}\bigg\}.
\end{align}
\begin{align}
\label{def:3x4}
&\frac{\td \sigma_{3\times 4}^T}{\td^2 \bp\, \td y_1} = 
\frac{e^2 g^2 Q^2 N_c}{(2\pi)^{8}}
\int^{\bz_1}_0\frac{\td z}{z} z_1^2 (\bz_1-z)
\int \td^{8}\bx  \left[\frac{K_1(QX)K_1(QX^\p)}{XX^\p}\right]_{\bx_2^\p = \bx_2} 
e^{i\bp\cdot\bx_{1^\p 1}} \nonumber \\
& \times
\Big[C_F \Big(S_{11^\p} - S_{13} S_{32} - S_{31^\p} S_{23} + 1\Big) 
- \frac{1}{2 N_c} \Big(S_{13} S_{32} + S_{31^\p} S_{23} - S_{12} - S_{21^\p}\Big)\Big]
\nonumber \\
&  \times4\Re
\bigg\{\bigg[\Big(z_1^2+(\bz_1-z)^2\Big)\frac{(\bx_{31}\de)(\bx_{32}\des)}{\bx_{31}^2\bx_{3 2}^2}
+ 
\frac{z_1(\bz_1-z)}{\bz_1(z_1+z)}\Big(\bz_1^2+(z_1+z)^2\Big) \frac{(\bx_{31}\des)(\bx_{3 2}\de)}{\bx_{31}^2\bx_{3 2}^2}
\bigg]
\nonumber\\
& \hspace{1.5cm}\times 
\Big[\big(z_1\bx_{12}+z\bx_{32}\big)\de\Big]\Big[\Big((\bz_1-z)\bx_{21^\p}+z\bx_{3 1^\p}\Big)\des\Big] \nonumber \\
% &+ \frac{z_1z_2}{(1-z_1)(1-z_2)}[(1-z_1)^2+(1-z_2)^2] \frac{(\bx_{31}\des)(\bx_{3 2}\de)}{\bx_{31}^2\bx_{3 2}^2}[(z_1\bx_{12}+z\bx_{32})\de][(z_2\bx_{2 1^\p}+z\bx_{3 1^\p})\des] \nonumber \\
&\hspace{1.5cm}
-\frac{z(\bz_1-z)(z_1+z)}{2\bz_1} \frac{(\bx_{31}\des)}{\bx_{31}^2}\Big[(z_1\bx_{12}+z\bx_{32})\de\Big]
-\frac{z_1z\bz_1}{2(z_1+z)}\frac{(\bx_{3 2}\de)}{\bx_{3 2}^2} 
\Big[\Big((\bz_1-z)\bx_{2 1^\p}+z\bx_{3 1^\p}\Big)\des\Big]\bigg\}.
\end{align}
\begin{align}
\label{def:1x3}
&\frac{\td \sigma_{1\times 3}^T}{\td^2 \bp\, \td y_1} =  - 
\frac{e^2 g^2 Q^2 N_c}{(2\pi)^{8}}
\int^{\bz_1}_0 \frac{\td z}{z} (\bz_1-z) \sqrt{(\bz_1-z)(z_1+z)}
\int\td^{8}\bx  \left[\frac{K_1(|\bx_{12}|Q_{1z}) K_1(QX^\p)}{X^\p}\right]_{\bx_2^\p = \bx_2} 
e^{i\bp\cdot\left(\bx_{1^\p 1} + \frac{z}{z_1}\bx_{31}\right)}
\nonumber \\
& \times
\Big[ C_F \Big( S_{13} S_{31^\p} - S_{31^\p} S_{23} - S_{12} + 1\Big)  
- \frac{1}{2 N_c} \Big(S_{11^\p} - S_{21^\p} - S_{13} S_{31^\p} + S_{31^\p} S_{23}\Big) \Big] \\
& \times  
4\Re \bigg\{(z_1+z)\Big(z_1^2+(\bz_1-z)^2\Big)\frac{(\bx_{12}\de)(\bx_{3 1^\p}\des)}{|\bx_{12}|\bx_{3 1^\p}^2} 
\frac{(\bx_{31}\de)\big[(z_1\bx_{1^\p 2}+z\bx_{3 2})\des\big]}{\bx_{31}^2} \nonumber \\
&+\Big((z_1+z)^3+\frac{z_1^2(\bz_1-z)^2}{z_1+z}\Big) \frac{(\bx_{12}\de)(\bx_{3 1^\p}\de)}{|\bx_{12}|\bx_{3 1^\p}^2} \frac{(\bx_{31}\des)\big[(z_1\bx_{1^\p 2}+z\bx_{3 2})\des\big]}{\bx_{31}^2} 
-\frac{z_1^2 (\bz_1-z) z}{2(z_1+z)} \frac{(\bx_{12}\de)}{|\bx_{12}|} 
\frac{(\bx_{31}\des)}{\bx_{31}^2}\bigg\}. \nonumber
\end{align}
\begin{align}
\label{def:1x4}
&\frac{\td \sigma_{1\times 4}^T}{\td^2 \bp\, \td y_1} =  - 
\frac{e^2 g^2 Q^2 N_c}{(2\pi)^{8}} 
\int ^{\bz_1}_0\frac{ \td z}{z} z_1 \sqrt{(\bz_1-z)(z_1+z)} 
\int\td^{8}\bx  \left[\frac{K_1(|\bx_{12}|Q_{1z}) K_1(QX^\p)}{X^\p}\right]_{\bx_2^\p = \bx_2}   
e^{i\bp\cdot\left(\bx_{1^\p 1} + \frac{z}{z_1} \cdot\bx_{31}\right)}
\nonumber \\
& \times
\Big[ C_F \Big(S_{13} S_{31^\p} - S_{31^\p} S_{23} - S_{12} + 1\Big)
- \frac{1}{2 N_c} \Big(- S_{13} S_{31^\p} + S_{31^\p} S_{23} + S_{11^\p} - S_{21^\p}\Big) \Big]
\nonumber \\ & \times 
4\Re \bigg\{ (z_1+z)\Big(z_1^2+(\bz_1-z)^2\Big)\frac{(\bx_{12}\de)(\bx_{3 2}\des)}{|\bx_{12}|\bx_{3 2}^2} 
\frac{(\bx_{31}\de) \big[\big((\bz_1-z)\bx_{2 1^\p}+z \bx_{3 1^\p}\big)\des\big]}{\bx_{31}^2} 
\nonumber \\
& \hspace{1.2cm}
+\frac{z_1(\bz_1-z)}{\bz_1}\Big(\bz_1^2+(z_1+z)^2\Big)\frac{(\bx_{12}\de)(\bx_{3 2}\de)}{|\bx_{12}|\bx_{3 2}^2} \frac{(\bx_{31}\des) \big[\big((\bz_1-z)\bx_{21^\p}+z \bx_{3 1^\p}\big)\des\big]}{\bx_{31}^2} 
\nonumber \\
&  \hspace{1.2cm}
-\frac{(\bz_1-z)z(z_1+z)^2}{2\bz_1} \frac{(\bx_{12}\de)}{|\bx_{12}|} 
\frac{(\bx_{31}\des)}{\bx_{31}^2}\bigg\}.
\end{align}
Before we proceed further, a few comments are in order. The shorthand notation $Q_{1z}$ used in the arguments of the Bessel functions in Eqs. \eqref{def:1x1},\eqref{def:1x2}\eqref{def:1x3} and \eqref{def:1x4} is given by 
\begin{align}
Q_{1z} \equiv Q \sqrt{(\bz_1-z) (z_1 + z)}\, . 
\end{align}
The transverse photon polarization vector $\be$ that appears in all NLO contributions is given by  
\begin{align}
%{\bf  \epsilon}_\perp 
\be 
= \frac{1}{\sqrt{2}}(1,i)
\end{align}%
and the radiation kernel $\Delta^{(3)}_{ij}$ used in Eq. \eqref{def:1x1} is defined as   
\begin{align}
\Delta^{(3)}_{ij} \equiv\frac{\bx_{3i}\cdot\bx_{3j}}{\bx_{3i}^2\bx_{3j}^2}.
\end{align}
The shorthand notation  $\bx_{2'}$ introduced in Eq. \eqref{def:1x2} is $\bx_{2^\p} = \bx_2 - \frac{z}{\bz_1} \bx_{23}$ so that it satisfies $\bx_{2^\p3} = \frac{(\bz_1-z)}{\bz_1}\bx_{23}$. Finally, the notations $X$ and $X'$ used in Eqs. \eqref{def:3x3}, \eqref{def:4x4}, \eqref{def:3x4}, \eqref{def:1x3} and \eqref{def:1x4} are defined as 
\begin{align}
\label{def:X}
X &= \sqrt{z_1 (\bz_1-z) \, \bx_{12}^2 + z_1  z \, \bx_{13}^2 + (\bz_1-z) z \, \bx_{23}^2} \\
\label{def:Xprime}
X' &= \sqrt{z_1 (\bz_1-z) \, \bx_{1'2'}^2 + z_1  z \, \bx_{1'3}^2 + (\bz_1-z) z \, \bx_{2'3}^2} 
\end{align}
Apart from the above listed contributions, NLO single inclusive quark production cross section receives real corrections from diagrams $\sigma_{2\times1}, \sigma_{4\times3}, \sigma_{3\times1}$ and $\sigma_{4\times1}$ that can be obtained by taking the complex conjugate of Eqs. \eqref{def:1x2}, \eqref{def:3x4}, \eqref{def:1x3} and \eqref{def:1x4} respectively. 

%%%%%%%%%%%
%%%%%%%%%%%

\subsubsection{Virtual corrections to single inclusive quark production at NLO}

In this subsection, we list the virtual corrections to the single inclusive quark production cross section at NLO that are obtained from the virtual corrections to double inclusive quark-antiquark cross section at NLO in DIS via transverse photon upon integration over the antiquark momenta $(z_2,\bq)$. The results read
\begin{align}
\label{def:sigma_5}
\frac{\td \sigma^{T}_5}{\td^2 \bp \, \td y_1} &=  
\frac{e^2 g^2 Q^2 N_c}{(2\pi)^{8} } 
\int_0^{z_1} \frac{\td z}{z} \bz_1 \sqrt{z_1 \bz_1} 
\int\td^{8} \bx 
\frac{K_1(QX_5) K_1(|\bx_{1^\p 2}|Q_1)}{X_5 |\bx_{1^\p 2}| \bx_{31}^2} \, 
e^{i\bp\cdot\left(\bx_{1^\p1} - \frac{z}{z_1} \bx_{31}\right)}
\nonumber \\
& \times
\Big[C_F \Big(S_{13} S_{31^\p}  - S_{32} S_{13} - S_{21^\p} + 1\Big)  
- \frac{1}{2 N_c} \Big(- S_{13} S_{31^\p}  + S_{32} S_{13} + S_{11^\p} - S_{12}\Big)\Big]
 \\
& \times 
\bx_{1^\p 2}\cdot\Big\{(z_1^2+\bz_1^2)\big[z_1^2+(z_1-z)^2\big]\bx_{32} 
+(z_1-z)\Big[z_1\big[z_1^2+(z_1-z)^2\big]+\bz_1\big[z_1\bz_1+(z_1-z)(\bz_1+z)\big]\Big]\bx_{13}\Big\}.
\nonumber
\end{align}
\begin{align}
\label{def:sigma_7}
\frac{\td \sigma^{T}_{7}}{\td^2 \bp \, \td y_1} & = 
\frac{e^2 g^2 Q^2 N_c}{(2\pi)^{8}} 
\int_0^{z_1}\frac{\td z}{z} z_1 (z_1-z) \sqrt{z_1 \bz_1}
\int\td^{8} \bx\,  
\frac{ K_1(QX_5)  K_1(|\bx_{1^\p2}|Q_1)}{X_5 |\bx_{1^\p 2}| \bx_{31}^2} 
e^{i\bp\cdot\left(\bx_{1^\p1} - \frac{z}{z_1} \bx_{31}\right)} 
\nonumber \\
& \times
\Big[C_F \Big(S_{13} S_{31^\p} - S_{32} S_{13} - S_{21^\p} + 1\Big)  
- \frac{1}{2 N_c} \Big(- S_{13} S_{31^\p} + S_{32} S_{13} + S_{11^\p} - S_{12}\Big) \Big]\nonumber \\
& \times \Bigg\{
\frac{4\Re}{\bx_{32}^2}\bigg\{ (\bx_{1^\p 2}\des)
\Big[\Big(\bx_{31}+\frac{\bz_1}{\bz_1+z}\bx_{23}\Big)\de\Big]
\Big[\frac{\bz_1 (z_1-z)}{z_1}\Big(z_1^2+(\bz_1+z)^2\Big)(\bx_{31}\de)(\bx_{32}\des) 
\nonumber \\
& \hspace{1.7cm}
+ (\bz_1+z)\Big(\bz_1^2+(z_1-z)^2\Big)(\bx_{32}\de)(\bx_{31}\des)\Big]\bigg\} 
-\frac{z_1 \bz_1z}{\bz_1+z}\bx_{31}\cdot\bx_{1^\p 2}\Bigg\}
\end{align}
\begin{align}
\label{def:sigma_9}
\frac{\td \sigma_9^{T}}{\td^2 \bp \, \td y_1} = - 
&\frac{e^2 g^2 Q^2 N_c}{2(2\pi)^6} 
\int_0^{z_1} \frac{\td z}{z}
\bz_1 (z_1^2+\bz_1^2) \Big[z_1^2 + (z_1-z)^2\Big] 
\int \td^6 \bx \frac{\bx_{12}\cdot\bx_{1^\p 2}}{|\bx_{12}||\bx_{1^\p 2}|}
K_1(|\bx_{12}|Q_1) K_1(|\bx_{1^\p 2}|Q_1)   
e^{i\bp\cdot\bx_{1^\p1}}
\nonumber \\ 
& \times
\Big[C_F \Big(S_{11^\p} - S_{12} - S_{21^\p} + 1\Big) \Big]
\int \frac{\td^2{\bk}}{(2\pi)^2} \frac{1}{\left(\bk-\frac{z}{z_1}\bp\right)^2}.
\end{align}
\begin{align}
\label{def:sigma_11}
\frac{\td \sigma_{11}^{T}}{\td^2 \bp\, \td y_1} &= 
i \frac{e^2 g^2 Q N_c}{(2\pi)^5} 
\int_0^{z_1} \frac{\td z}{z}  \sqrt{z_1 \bz_1}(z_1^2+\bz_1^2) 
\frac{[(z_1-z)^2+z_1^2]}{(z_1-z)} 
\int \td^6 \bx
\frac{K_1(|\bx_{1^\p2}|Q_1)}{|\bx_{1^\p2}|} \,
e^{i\bp\cdot\bx_{1^\p1}}
\nonumber \\
& \times
\Big[C_F \Big(S_{11^\p} - S_{12} - S_{21^\p} + 1\Big) \Big]
\int \frac{\td^2{\bk_2}}{(2\pi)^2} \int \frac{\td^2{\bk_1}}{(2\pi)^2}  \frac{\bk_1\cdot\bx_{1^\p2} \,}
{\big[\bk_1^2+Q_1^2\big]\Big[z Q^2 +\frac{z \bk_1^2}{z_1\bz_1} +\frac{z_1}{z_1-z}\bk_2^2\Big]} \, e^{i\bk_1\cdot\bx_{12}}
\end{align}
\begin{align}
\frac{\td \sigma_{12}^{T}}{\td^2 \bp\, \td y_1} &= - 
i \frac{e^2 g^2 Q N_c}{(2\pi)^5}  
\int_0^{\bz_1} \frac{\td z}{z} \sqrt{\frac{z_1}{\bz_1}} z_1 (z_1^2+\bz_1^2) 
\frac{[(\bz_1-z)^2+\bz_1^2]}{(\bz_1-z)} 
\int \td^6 \bx 
\frac{K_1(|\bx_{1^\p2}|Q_1)}{|\bx_{1^\p2}|} 
e^{i\bp\cdot\bx_{1^\p1}} 
\nonumber\\
& \times
\Big[C_F \Big(S_{11^\p} - S_{12} - S_{21^\p} + 1\Big)\Big]
\int \frac{\td^2{\bk_2}}{(2\pi)^2} \int \frac{\td^2{\bk_1}}{(2\pi)^2}  \frac{\bk_1\cdot\bx_{1^\p 2}}
{\big[\bk_1^2+Q_1^2\big]\Big[z Q^2 +\frac{z \bk_1^2}{z_1\bz_1} +\frac{\bz_1}{\bz_1-z}\bk_2^2\Big]}\, e^{i\bk_1\cdot\bx_{21}}
\end{align}
\begin{align}
\label{def:sigma_13-1}
&\frac{\td \sigma_{13(1)}^{T}}{\td^2 \bp\, \td y_1} =  
\frac{e^2 g^2 Q^2 N_c}{(2\pi)^6}
\int_{0}^{\bz_1} \td z \, z_1 \sqrt{z_1 \bz_1 (z_1+z)(\bz_1-z)} 
\\
&\times
\int \td^8\bx\,
\Big[ C_F \Big(S_{2^\p1^\p} S_{12} - S_{12} - S_{2^\p1^\p} + 1\big)  
- \frac{1}{2 N_c} \Big(- S_{2^\p1^\p} S_{12} + S_{122^\p1^\p}\Big) \Big]
\int \frac{\td^2{\bq}}{(2\pi)^2} \, e^{i\bp\cdot\bx_{1^\p 1}} \, e^{i\bq\cdot\bx_{2^\p 2}}
\int \frac{\td^2{\bk}}{(2\pi)^2} \, e^{i\bk\cdot\bx_{21}}
\nonumber \\ 
& \times
\frac{1}{|\bx_{12}|} K_1\Big(|\bx_{12}|Q\sqrt{(z_1+z)(\bz_1-z)}\Big)\; 
\frac{1}{|\bx_{1^\p 2^\p}|} K_1\big(|\bx_{1^\p 2^\p}|Q_1\big) \; 
4\Re\bigg\{ \frac{(\bx_{12}\de)(\bx_{1^\p 2^\p}\des)}{(\bz_1\bk-z\bq)^2}
\nonumber\\
&\times
\bigg[
\frac{\bz_1(\bz_1-z)\big[z_1(z_1+z)+\bz_1(\bz_1-z)\big]}{2z} 
+ \frac{1}{\Big[\frac{(z_1\bk-z\bp)^2}{z_1(z_1+z)}-\frac{(\bz_1\bk-z\bq)^2}{\bz_1 (\bz_1-z)}\Big]}
\Big(f_1(z,z_1;\bk,\bp,\bq;{\bf \epsilon},{\bf \epsilon^*}) +f_2(z,z_1;\bk,\bp,\bq;{\bf \epsilon},{\bf \epsilon^*})\Big)\bigg]\bigg\} \nonumber
\end{align}
\begin{align}
\label{def:sigma_13-2}
&\frac{\td \sigma_{13(2)}^{T}}{\td^2 \bp\, \td y_1} =  
\frac{e^2 g^2 Q^2 N_c}{(2\pi)^6}
\int_{0}^{z_1} \td z\,  z_1 \sqrt{z_1 \bz_1 (z_1-z)(\bz_1+z)}
\\
& \times
\int \td^8\bx\,
\Big[ C_F \Big(S_{2^\p1^\p} S_{12} - S_{12} - S_{2^\p1^\p} + 1\Big)  
- \frac{1}{2 N_c} \Big(- S_{2^\p1^\p} S_{12} + S_{122^\p1^\p}\Big)\Big] 
\int \frac{\td^2{\bq}}{(2\pi)^2}\, e^{i\bp\cdot\bx_{1^\p 1}} \, e^{i\bq\cdot\bx_{2^\p 2}}
\int \frac{\td^2{\bk}}{(2\pi)^2} \, e^{i\bk\cdot\bx_{12}} 
\nonumber \\
& \times 
\frac{1}{|\bx_{12}|} K_1\Big(|\bx_{12}|Q\sqrt{(z_1-z)(\bz_1+z)}\Big) \, 
\frac{1}{|\bx_{1^\p 2^\p}|} K_1\big(|\bx_{1^\p 2^\p}|Q_1\big) \, 
4\Re\bigg\{  \frac{(\bx_{12}\de)(\bx_{1^\p 2^\p}\des)}{(z_1\bk-z\bp)^2}
\nonumber\\
&\times 
\bigg[ 
\frac{z_1(z_1-z)\big[z_1(z_1-z)+\bz_1(\bz_1+z)\big]}{2z} 
%\nonumber \\
%&
+\frac{1}{\Big[\frac{(z_1\bk-z\bp)^2}{z_1(z_1-z)}-\frac{(\bz_1\bk-z\bq)^2}{\bz_1 (\bz_1+z)}\Big]}
\Big(f_3(z,z_1;\bk,\bp,\bq;{\bf \epsilon},{\bf \epsilon^*}) +f_4(z,z_1;\bk,\bp,\bq;{\bf \epsilon},{\bf \epsilon^*})\Big)\bigg]\bigg\} \nonumber
\end{align}
\begin{align}
\label{def:sigma_14-1}
\frac{ \td \sigma_{14(1)}^{T}}{\td^2 \bp \, \td y_1} &= - i 
\frac{e^2 g^2 Q N_c}{(2\pi)^5} 
\int_0^{z_1} \frac{\td z}{z} z_1 \sqrt{z_1 \bz_1} 
\int \td^6 \bx\;  
\Big[C_F \Big(S_{11^\p} - S_{12} - S_{21^\p} + 1\Big)\Big] \, 
\frac{1}{|\bx_{1^\p2}|} \, K_1\big(|\bx_{1^\p2}|Q_1\big) \,  
e^{i\bp\cdot\bx_{1^\p 1}}
\nonumber \\
& \times 
\int  \frac{\td^2{\bk_1}}{(2\pi)^2} \, \int \frac{\td^2{\bk_2}}{(2\pi)^2}\, e^{i\bk_2\cdot\bx_{12}}
\frac{1}{\big[\big(\bk_1-\frac{z_1-z}{z_1}\bk_2\big)^2+
\frac{z(z_1-z)}{\bz_1 z_1^2} \bk_2^2 + \frac{z}{z_1}(z_1-z)Q^2\big]}
\nonumber\\
& \times
\bigg\{ 
\frac{\big[z_1(z_1-z)+\bz_1(\bz_1+z)-z(1-z)\big] (\bk_2\cdot\bx_{1^\p 2})} {\big[\bk_2^2 + Q_1^2\big]}
+\frac{(z_1-z)\big[1+z^2-2\bz_1(z_1-z)\big](\bk_1\cdot\bx_{1^\p 2}) }
{z_1\big[\bk_1^2 + (z_1-z)(\bz_1+z)Q^2\big] } 
\nonumber \\
& \hspace{0.7cm}
-\frac{Q^2 (z_1-z)\br{2z_1\bz_1z \, (\bk_1\cdot\bx_{1^\p2}) 
+ z(z+\bz_1-z_1)^2(\bk_2\cdot\bx_{1^\p2})}}
{z_1\big[\bk_1^2 + (z_1-z)(\bz_1+z)Q^2\big]\big[\bk_2^2 + Q_1^2\big]
}
\bigg\}
\end{align}
\begin{align}
\label{def:sigma_14-2}
\frac{ \td \sigma_{14(2)}^{T}}{\td^2 \bp \, \td y_1} &= - i 
\frac{e^2 g^2 Q N_c}{(2\pi)^5}  
\int_0^{\bz_1} \frac{\td z}{z} z_1 \sqrt{z_1 \bz_1} 
\int \td^6 \bx \, \Big[ C_F 
\Big(S_{11^\p} - S_{12} - S_{21^\p} + 1\Big)\Big] \, 
\frac{1}{|\bx_{1^\p 2}|} \, K_1\big(|\bx_{1^\p 2}|Q_1\big) \, 
e^{i\bp\cdot\bx_{1^\p 1}} 
\nonumber \\
&\int  \frac{\td^2{\bk_1}}{(2\pi)^2} \int \frac{\td^2{\bk_2}}{(2\pi)^2} \, e^{i\bk_2\cdot\bx_{12}} 
\frac{1}{\big[\big(\bk_1-\frac{\bz_1-z}{\bz_1}\bk_2\big)^2
+\frac{z(\bz_1-z)}{z_1 \bz_1^2} \bk_2^2 + \frac{z}{\bz_1}(\bz_1-z)Q^2\big]}
\nonumber\\
& \times 
\bigg\{ 
\frac{\big[\bz_1(\bz_1-z)+z_1(z_1+z)-z(1-z)\big] (\bk_2\cdot\bx_{1^\p2})}
{\big[\bk_2^2 + Q_1^2\big] }
+\frac{(\bz_1-z)(1+z^2-2z_1(\bz_1-z))(\bk_1\cdot\bx_{1^\p 2}) }
{\bz_1\big[\bk_1^2 + (\bz_1-z)(z_1+z)Q^2\big]} \nonumber \\
&\hspace{0.7cm}
-\frac{Q^2(\bz_1-z)\big[2z_1\bz_1z\, (\bk_1\cdot\bx_{1^\p 2}) 
+ z(z+z_1-\bz_1)^2(\bk_2\cdot\bx_{1^\p 2})\big]}{\bz_1\big[\bk_1^2 + (\bz_1-z)(z_1+z)Q^2\big]
\big[\bk_2^2 + Q_1^2\big]}
\bigg\}
\end{align}
In Eqs. \eqref{def:sigma_5} and \eqref{def:sigma_7} we have used $X_5$ which is defined as 
\begin{align}
\label{def:X_5}
X_5 &= \sqrt{\bz_1(z_1-z)\bx_{12}^2 + z(z_1-z) \bx_{13}^2 + \bz_1 z\, \bx_{23}^2}, 
\end{align}
Moreover, for convenience the results given in Eqs. \eqref{def:sigma_13-1} and \eqref{def:sigma_13-2} are written in terms of functions $f_i(z,z_1;\bk,\bp,\bq;{\bf \epsilon},{\bf \epsilon^*})$  whose explicit expressions read 
\begin{align}
f_1(z,z_1;\bk,\bp,\bq;{\bf \epsilon},{\bf \epsilon^*})&= z_1\bz_1 z
\bigg[
z_1z\, \Big(\frac{\bp\de}{z_1}-\frac{\bq\de}{\bz_1}\Big)
\Big(\frac{\bp\des}{z_1}-\frac{\bk\des}{z}\Big) 
+z^2\, \Big(\frac{\bk\de}{z}-\frac{\bq\de}{\bz_1}\Big)
\Big(\frac{\bp\des}{z_1}-\frac{\bk\des}{z}\Big) 
\nonumber\\
&\hspace{1.5cm}
-\bz_1z \, \Big(\frac{\bk\de}{z}-\frac{\bq\de}{\bz_1}\Big)
\Big(\frac{\bp\des}{z_1}-\frac{\bq\des}{\bz_1}\Big) - p\cdot q\bigg]
\\
f_2(z,z_1;\bk,\bp,\bq;{\bf \epsilon},{\bf \epsilon^*})&= \frac{\bz_1^2 z (\bz_1-z)}{(z_1+z)}
\bigg [z_1z \, \Big(\frac{\bp\des}{z_1}-\frac{\bq\des}{\bz_1}\Big)
\Big(\frac{\bp\de}{z_1}-\frac{\bk\de}{z}\Big) +z^2\Big(\frac{\bk\des}{z}-\frac{\bq\des}{\bz_1}\Big)
\Big(\frac{\bp\de}{z_1}-\frac{\bk\de}{z}\Big) 
\nonumber\\
& \hspace{2cm}
-\bz_1z\, \Big(\frac{\bk\des}{z}-\frac{\bq\des}{\bz_1}\Big)
\Big(\frac{\bp\de}{z_1}-\frac{\bq\de}{\bz_1}\Big) - p\cdot q\bigg]
\end{align}
\begin{align}
f_3(z,z_1;\bk,\bp,\bq;{\bf \epsilon},{\bf \epsilon^*}) &= z_1\bz_1z
\bigg[ z_1z\, \Big(\frac{\bp\des}{z_1}-\frac{\bq\des}{\bz_1}\Big)
\Big(\frac{\bp\de}{z_1}-\frac{\bk\de}{z}\Big) 
-z^2 \, \Big(\frac{\bk\des}{z}-\frac{\bq\des}{\bz_1}\Big)\Big(\frac{\bp\de}{z_1}-\frac{\bk\de}{z}\Big) 
\nonumber\\
& \hspace{1.5cm}
- \bz_1z \,\Big(\frac{\bk\des}{z}-\frac{\bq\des}{\bz_1}\Big)
\Big(\frac{\bp\de}{z_1}-\frac{\bq\de}{\bz_1}\Big) + p\cdot q \bigg] \\
f_4(z,z_1;\bk,\bp,\bq;{\bf \epsilon},{\bf \epsilon^*}) &= 
\frac{z_1^2 z(z_1-z)}{\bz_1+z}\bigg[
z_1z \, \Big(\frac{\bp\de}{z_1}-\frac{\bq\de}{\bz_1}\Big)
\Big(\frac{\bp\des}{z_1}-\frac{\bk\des}{z}\Big) 
-z^2\, \Big(\frac{\bk\de}{z}-\frac{\bq\de}{\bz_1}\Big)
\Big(\frac{\bp\des}{z_1}-\frac{\bk\des}{z}\Big) 
\nonumber\\
& \hspace{2.5cm}
-\bz_1z \, \Big(\frac{\bk\de}{z}-\frac{\bq\de}{\bz_1}\Big)
\Big(\frac{\bp\des}{z_1}-\frac{\bq\des}{\bz_1}\Big) + p\cdot q\bigg]
\end{align}
Finally, the results presented in Eqs. \eqref{def:sigma_5}-\eqref{def:sigma_14-2} are the virtual contributions that are computed by considering the virtual diagrams in the amplitude and leading order diagrams in the complex conjugate amplitude. The virtual contributions that will come from considering the leading order diagrams in the amplitude and virtual diagrams in the complex conjugate amplitude, i.e. $\sigma^{\star}_5,\sigma^{\star}_7,\sigma^{\star}_9,\sigma^{\star}_{11},\sigma^{\star}_{12},\sigma^{\star}_{13(1)},\sigma^{\star}_{13(2)}, \sigma^{\star}_{14(1)}$ and $\sigma^{\star}_{14(2)}$ can be trivially obtained by taking the complex conjugate of the Eqs. \eqref{def:sigma_5}-\eqref{def:sigma_14-2}.

\section{Divergences}
\label{divergences}

In this section, we discuss the ultraviolet and infrared divergences that appear in the formal expressions above. We show that they either cancel or get absorbed into redefinitions of the relevant non-perturbative quantities (fragmentation function and dipole scattering amplitude), as expected.

\subsection{Ultraviolet divergences}

Ultraviolet (UV) divergences arise when the transverse momentum in the loop goes to infinity, or equivalently in coordinate space when the transverse coordinate of the radiated gluon $\bx_3$ approaches that of a quark ($\bx_1,\bx_1^\p$) or anti-quark ($\bx_2$). 

In the case of the double-inclusive cross-section, these divergences appeared only in virtual corrections, and therefore could be dealt with already at the amplitude level \cite{Caucal:2021ent,Taels:2022tza,Bergabo:2023wed}, where complete cancellation occurred. Things work a bit differently in the case of single-inclusive production as some of the diagrams involved in the UV cancellation prior to the anti-quark integration ($\sigma_{6}$ and $\sigma_{10}$) are no longer present; conversely, a UV divergence now appears in the real correction $\sigma_{4\times 4}$, as $\bx_3 \to \bx_2$. 

After combination with those appearing in virtual corrections (namely when $\bx_3 \to \bx_1$ in $\sigma_5$; $\bx_3 \to \bx_1^\p$ in $\sigma_5^{*}$; $|\bk|\to\infty$ in $\sigma_{9}$ and $\sigma_{9}^{*}$; $|\bk_2| \to \infty$ in $\sigma_{11}$, $\sigma_{11}^{*}$, $\sigma_{12}$ and $\sigma_{12}^{*}$; $|\bk_1| \to \infty$ in $\sigma_{14}$ and $\sigma_{14}^{*}$), we find that these UV divergences all cancel as follows:
\begin{align}
&\left[\dd \sigma_{5}+\dd \sigma_{11}\right]_{\text{UV}} = 
\left[\dd \sigma_{5}^{*}+\dd \sigma_{11}^{*}\right]_{\text{UV}} = 0,\nonumber \\
%&\left[\dd \sigma_{6}+\dd \sigma_{12} \right]_{\text{UV}} = 0, \nonumber \\
%&\left[\dd \sigma_{9}+\dd\sigma_{10}+\dd\sigma_{14(1)}+\dd\sigma_{14(2)}\right]_{\text{UV}}  = 0 \nonumber \\
&\left[2\ \mbox{Re}\left(\dd\sigma_{9} + \dd\sigma_{12} +  \dd\sigma_{14(1)} +  \dd\sigma_{14(2)}\right)  + 
\dd\sigma_{4\times 4} 
\right]_{\text{UV}}  = 0,
%&\left[\dd \sigma_{3\times 3}+\dd \sigma_{11}+\dd\sigma_{10}+\dd\sigma_{14(1)}+\dd\sigma_{14(2)}\right]_{\text{UV}}  = 
%\label{UV}
%&\left[\dd \sigma_{1\times1}+\dd\sigma_{3\times 3}+2\dd\sigma_{1\times 3}\right]_{\text{UV}} = 0
\label{UV}
\end{align}
where the cancellation in the last line above happens after integrating over $z$, in a similar fashion to the case of SIDIS with a longitudinal photon exchange\cite{Bergabo:2022zhe}.

\subsection{Collinear divergences and evolution of fragmentation functions}

Collinear divergences occur when the angle $\theta$ between the radiated gluon 3-momentum and the final-state quark 3-momentum goes to zero at finite $\bk$ and $z$. They are only present in $\sigma_{1\times 1}$, $\sigma_{9}$ and $\sigma_{9}^{*}$, where, in momentum space expressions, they appear in the denominator $1/(z_1\bk-z\bp)^2$ which behaves as $\sim1/\theta^2$ when $z_1\bk\to z\bp$. In coordinate space, due to the following expressions,
\begin{align}
\int \dd^2 \bk\ \frac{e^{i\bk\cdot\bx_{1^\p1}}}{\left(\bk-\frac{z}{z_1}\bp\right)^2}&=
e^{i\frac{z}{z_1}\bp\cdot\bx_{1^\p1}} 
\int \dd^2 \bx_3\ \Delta^{(3)}_{11^\p}\ ,\\
\int \dd^2 \bk\ \frac{1}{\left(\bk-\frac{z}{z_1}\bp\right)^2}&=\int \dd^2 \bx_3\ \frac{1}{\bx_3^2}\ ,
\end{align}
this limit translates to $|\bx_3|\to \infty$, but it should be kept in mind that those collinear divergences do not correspond to the gluon transverse momentum vanishing (they are no such singularities in our expressions, unless $z$ also vanishes, as we will see in the following section).

The collinear divergences do not cancel, and must be absorbed into a redefinition of the quark-to-hadron fragmentation function $D_{h/q}$. The procedure is identical to that in the case of di-hadrons and is detailed e.g. in~\cite{Bergabo:2022tcu}; here we shall highlight the main steps. First, one writes the hadronic cross section in terms of the partonic cross section as follows
\begin{align}
\frac{\dd \sigma^{\gamma^*A \to h (\bp_h,y_h)X}}{\dd^2 \bp_h\, \dd y_h} = \int_{z_h}^1 \frac{\td z_1}{z_h} \ D_{h/q}(z_h/z_1) \frac{\dd \sigma^{\gamma^*A \to q ((z_1/z_h)\bp_h,y_1)X}}{\dd^2 \bp\, \dd y_1}\ ,
\label{hadronic}
\end{align}
where $z_h$ and $\bp_h$ stand for the fraction of photon longitudinal momentum carried by the produced hadron and its transverse momentum, respectively. Then,
the divergent part in $\sigma_{1\times 1}+2\mbox{Re}(\sigma_{9})$ can be isolated with the help of a factorization scale $\mu$, it reads
\begin{align}
\dd\sigma_{LO}\times
\frac{\alpha_s C_F}{2\pi} 
\left( 
\int_0^{1 - z_h} \frac{\dd z}{z}\frac{1 + (1-z)^2}{1 - z} D_{h/q}(\frac{z_h}{1 - z})
- 
\int_0^1 \frac{\dd z}{z} [1 + (1-z)^2] D_{h/q}(z_h)  
\right)
\int_{1/\mu^2} \frac{\dd^2 \bx}{\bx^2} \ ,
\label{Pqq}\end{align}
where in that expression $\dd\sigma_{LO}$ means the leading-order hadron-level cross-section stripped of its fragmentation function $D_{h/q}(z_h)$. This divergent part is then absorbed into the leading-order cross section using a redefinition of the bare $D_{h/q}$ into a fragmentation function renormalized at the scale $\mu$:
\begin{align}
D_{h/q}(z, \mu^2) = D_{h/q}(z) + \mathcal{O}(\alpha_s\ln\mu^2)
\label{renorm}
\end{align}
where the $\mathcal{O}(\alpha_s)$ term in precisely the one in \eqref{Pqq}. What is left of $\sigma_{1\times 1}+2\mbox{Re}(\sigma_{9})$ is a finite piece that depends on $\mu$, while the fragmentation function in the LO term is obtained from \eqref{renorm} as the solution of the DGLAP evolution equation
$(\dd/\dd\ln\mu^2) D_{h/q}(z,\mu^2) = \dots$ where the right-hand side is now free of the infrared divergence. Finally, 
in the left-over finite piece of $\sigma_{1\times 1}+2\mbox{Re}(\sigma_{9})$, as well as in all the other NLO corrections, the bare fragmentation function can be replaced by the renormalized one, up to corrections of even higher order beyond the accuracy of our calculation.

\subsection{Soft divergences}

We refer to soft divergences as those that arise when both the transverse and longitudinal momenta in the loop go to zero simultaneously, i.e. $\bk\to\mathbf{0},z\to 0$. In our coordinate space expression, this translates to $|\bx_3|\to \infty,z\to 0$. Those fully cancel as follows:
\begin{align}
&\left[\dd \sigma_{1\times 1} + 2\,\mbox{Re}(\dd \sigma_{9}) \right]_{\text{soft}}= 0, \nonumber \\
&\left[\dd \sigma_{1\times 2} +  \dd\sigma_{13(1)} + \dd\sigma_{13(2)}\right]_{\text{soft}} = \left[\dd \sigma_{2\times 1} +  \dd\sigma_{13(1)}^{*} + \dd\sigma_{13(2)}^{*}\right]_{\text{soft}} = 0 ,\nonumber\\
&\left[\dd \sigma_{3\times 3} + \dd \sigma_{4\times 4} + 2\,\mbox{Re}(\dd \sigma_{3\times 4})\right]_{\text{soft}} = 0,\nonumber\\
%&\left[\dd \sigma_{1\times 3} + \dd \sigma_{1\times 4} \right]_{\text{soft}}= 0,\nonumber\\
&\left[\dd \sigma_{1\times 3} + \dd \sigma_{1\times 4}\right]_{\text{soft}} = \left[\dd \sigma_{3\times 1} + \dd \sigma_{4\times 1}\right]_{\text{soft}} = 0,\nonumber\\
&\left[\dd \sigma_{5}+\dd\sigma_{7} \right]_{\text{soft}}= \left[\dd \sigma^*_{5}+\dd\sigma^*_{7} \right]_{\text{soft}}= 0, \nonumber \\
%&\left[\dd \sigma_{6}+\dd\sigma_{8}\right]_{\text{soft}} = 0, \nonumber \\
&\left[\dd \sigma_{11} + \dd \sigma_{14(1)}\right]_{\text{soft}} = \left[\dd \sigma_{11}^* + \dd \sigma_{14(1)}^*\right]_{\text{soft}} = 0,  \nonumber \\
&\left[\dd \sigma_{12} + \dd \sigma_{14(2)}\right]_{\text{soft}} = \left[\dd \sigma_{12}^* + \dd \sigma_{14(2)}^*\right]_{\text{soft}} = 0 .
\end{align}

\subsection{Rapidity divergences and small $x$ evolution}

Rapidity divergences appear when $z \rightarrow 0$ at non-zero transverse momentum. In that case, none of the soft cancellations of Section IV.C survive,
except 
\begin{align}
\left[\dd \sigma_{1\times 2} +  \dd\sigma_{13(1)} + \dd\sigma_{13(2)}\right]_{z=0} = \left[\dd \sigma_{2\times 1} +  \dd\sigma_{13(1)}^{*} + \dd\sigma_{13(2)}^{*}\right]_{z=0} = 0.
\end{align}
We note that when considering the double-inclusive cross-section, this was not the case: $\sigma_{1\times 2}$, $\sigma_{13}$, $\sigma_{2\times 1}$ and $\sigma_{13}^{*}$
also had to be considered in what follows.

The surviving rapidity divergences can be isolated by introducing a rapidity factorization scale $z_f$
and writing the $z$ integral as     
\begin{align}
\int_0^1 \frac{\dd z}{z} f (z) =  \left\{\int_0^{z_f} \frac{\dd z}{z}  + 
\int_{z_f}^1 \frac{\dd z}{z} \, \right\} \, f (z).
\end{align}
where the first integral contains the $z\to0$ divergence while the second integral contains the finite NLO pieces. 
When putting together all the divergent terms, we obtain 
\begin{eqnarray}
\left.\frac{\dd\sigma^{T}_{NLO}}{d^{2} \boldsymbol{p} dy_1}\right|_{z\to0}\!\!\! && = \frac{e^{2}Q^{2}N_{c}}{(2\pi)^{5}}(1-z_{1})z_{1}^{2}[z_{1}^{2}+(1-z_{1})^{2}]\int d^{6}\bx K_{1}(|\bx_{12}|Q_{1})K_{1}(|\bx_{1'2}|Q_{1})\frac{\bx_{12}\cdot\bx_{1'2}}{\bx_{12}^{2}\bx_{1'2}^{2}}e^{i\bp\cdot\bx_{1'1}}\nonumber \\
 && \times\frac{\alpha_{s}N_c}{2\pi^{2}}\!\int_{0}^{z_f}\frac{dz}{z}\int d^{2}\bx_{3}\left[\frac{\bx_{11'}^{2}}{\bx_{13}^{2}\bx_{1'3}^{2}}\left(S_{31'}S_{13}\!-\!S_{11'}\right)\!-\!\frac{\bx_{12}^{2}}{\bx_{13}^{2}\bx_{23}^{2}}\left(S_{31}S_{23}\!-\!S_{12}\right)\!-\!\frac{\bx_{21'}^{2}}{\bx_{23}^{2}\bx_{1'3}^{2}}\left(S_{31'}S_{23}\!-\!S_{21'}\right)\right]\ ,
\label{BK}
\end{eqnarray}
similarly to the case of SIDIS with a longitudinal photon exchange\cite{Bergabo:2022zhe}, although we have now obtained that result at finite $N_c$.

The individual contributions of all the diagrams to \eqref{BK} are listed in Appendix B. As for contributions of diagrams  $\sigma_{1\times 1}$, $\sigma_{9}$ and $\sigma_{9}^{*}$, their $z\to0$ limit have been obtained prior to the extraction of (\ref{Pqq}). Because the collinear divergence in (\ref{Pqq}) cancels in the $z\to0$ limit, their contribution to \eqref{BK} may also be extracted from $[\sigma_{1\times 1}+2\mbox{Re}(\sigma_{9})]_{finite}$, while ignoring the $\mu$ dependence, providing $z_f$ is small enough. Behind the question of how small $z_f$ should be lies the issue of large Sudakov logarithms discussed in \cite{Altinoluk:2024vgg}.

Comparing \eqref{BK} to the LO result in Eq. \ref{LOdsig-sidis-quark}
it is clear that the terms inside the square bracket correspond to the BK/JIMWLK evolution~
\cite{Balitsky:1995ub,Kovchegov:1999yj,Jalilian-Marian:1997qno,Jalilian-Marian:1997jhx,Jalilian-Marian:1997ubg} 
of the dipoles that appear in the LO cross section. In other words, this rapidity-divergent contribution can be absorbed into the LO cross section by renormalizing the dipole amplitudes at the scale $z_f$:
\begin{align}
S_{ij}(z_f) = S_{ij} + \mathcal{O}(\alpha_s\ln(z_f))
\end{align}
and obtaining them as the solution of the BK equation $\dd/\dd\ln(z_f)S_{ij}(z_f)=\dots$ where the right-hand side is now divergence-free.

The contribution of the 
$\int_{z_f}^1\dd z$ region is finite and $z_f$ dependent. Along with the other NLO corrections not affected by the rapidity divergence, they constitute the NLO correction to the LO result. Up to higher-order corrections beyond the accuracy of our calculation, we are free to promote all the bare dipole amplitudes involved to renormalized ones. The full result for the single inclusive cross section at next-to-leading order can be written schematically as
\begin{align}
\dd \sigma^{\gamma^* A \to h X} = \dd \sigma_{LO}[{S_{ij}(z_f)},D_{h/q}(z_h,\mu^2)] + \dd \sigma_{NLO}^{\text{finite}}(z_f,\mu^2).
\end{align}

\section{Summary and outlook}
\label{conc}

We have calculated the next to leading order (NLO) corrections to single inclusive hadron production in DIS (SIDIS) at small $x$ in the color glass condensate formalism when the exchanged photon is transversely polarized. 
We have shown that all UV and soft divergences cancel. There are collinear divergences that are absorbed into quark-hadron fragmentation function and cause it to evolve with DGLAP evolution equation. On the other hand the rapidity divergences are absorbed into rapidity evolution of dipoles with BK equation. The final result is finite and given at finite $N_c$.  These expressions can be used for phenomenological studies of CGC at the proposed Electron-Ion Collider (EIC), as well as in the high energy Ultraperipheral heavy ion collisions at RHIC and the LHC. One can investigate the rapidity dependence of production cross section at fixed values of hadron transverse momentum which should be a good proxy for confirming the predictions of CGC for $x_g$ dependence of observables. Furthermore, one can vary the transverse momentum of produced hadrons at fixed rapidity in order to explore the dense saturation region as well as the kinematics where the target is dilute but is affected by saturation boundary (geometric scaling region).   

%{\bf mention impact parameter, $A$ dependence,...?}

SIDIS has the advantage that one can reach a smaller value of $x_g$, fraction of hadron momentum carried by a gluon, than double inclusive hadron production and therefore will be a better process in which to investigate gluon saturation effects. Furthermore, in the back to back kinematic regime of double inclusive hadron production one is necessarily in the kinematics where there is a hierarchy of transverse momenta/virtuality which would necessitate a resummation of Sudakov effects. Seemingly this may be largely avoided in SIDIS if one chooses to work in the kinematics where the produced hadron transverse momentum and photon virtuality are comparable. In this kinematics saturation effects can be investigated without having to worry about Sudakov effects. Nevertheless due to the large phase space available at EIC it is desirable to include effects even in SIDIS. In \cite{Altinoluk:2024vgg} Sudakov double logarithms are extracted from the NLO SIDIS cross section in the large virtuality limit of the incoming transverse photon. A natural continuation of this study which we are planing to address is the extraction of the single Sudakov logarithms in SIDIS but this is beyond the scope of current manuscript and is left for future work.

\acknowledgements

JJM thanks CPHT, Ecole Polytechnique for hospitality during the visit when this work was finalized. TA is supported in part by the National Science Centre (Poland) under the research Grant No. 2023/50/E/ST2/00133
(SONATA BIS 13).  JJM is supported by the US DOE Office of Nuclear Physics through Grant No. DE-
SC0002307. This material is based upon
work supported by the U.S. Department of Energy, Office of Science, Office of Nuclear Physics, within the framework
of the Saturated Glue (SURGE) Topical Theory Collaboration.  

\appendix

\section{Quark-antiquark production cross section at NLO}
\label{app:dihadron_NLO}

%{\bf this Appendix is corrected: $3^\p = 3$ in wilson lines,... }

%\vspace{0.3in}

Here we write the quark antiquark production cross section including the full $N_c$ structure,

{\small 
\begin{align}
&\frac{\dd \sigma_{1\times 1}^T}{\dd^2 \bp \, \dd^2 \bq\, \dd y_1 \, \dd y_2} = 
\frac{e^2 g^2 Q^2 N_c\, z_2^2 (1-z_2) [z_1^2 z_2^2 + (z_1^2+z_2^2)(1-z_2)^2+ (1-z_2)^4]}{(2\pi)^{10}z_1}  
\int \dd^{10}\bx \, C_F \left\{ S_{122^\p1^\p} - S_{12} - S_{2^\p1^\p} + 1\right\}
\nonumber \\
& 
\int \frac{\dd z}{z} 
e^{i\bp\cdot\bx_{1^\p 1}} e^{i\bq\cdot\bx_{2^\p 2}} 
e^{i\frac{z}{z_1}\bp\cdot\bx_{1^\p 1}}
K_1(|\bx_{12}|Q_2)K_1(|\bx_{1^\p 2^\p}|Q_2) \frac{\bx_{12}\cdot\bx_{1^\p 2^\p}}{|\bx_{12}||\bx_{1^\p 2^\p}|}\, 
\Delta^{(3)}_{1^\p 1} . \\
%%%%%%%
\\
%%%%%%%
&\frac{\dd \sigma_{2\times 2}^T}{\dd^2 \bp \, \dd^2 \bq\, \dd y_1 \, \dd y_2} = 
\frac{e^2 g^2 Q^2 N_c \, z_1^2 (1-z_1) [z_1^2 z_2^2 + (z_1^2+z_2^2)(1-z_1)^2+ (1-z_1)^4]}{(2\pi)^{10}z_2} 
\int \dd^{10}\bx \, 
C_F \left\{ S_{122^\p1^\p} - S_{12} - S_{2^\p1^\p} + 1\right\} 
\nonumber \\
& 
\int\frac{\dd z}{z} 
e^{i\bp\cdot\bx_{1^\p 1}} e^{i\bq\cdot\bx_{2^\p 2}} 
e^{i\frac{z}{z_2}\bq\cdot\bx_{2^\p 2}} 
K_1(|\bx_{12}|Q_1)K_1(|\bx_{1^\p 2^\p}|Q_1) 
\frac{\bx_{12}\cdot\bx_{1^\p 2^\p}}{|\bx_{12}||\bx_{1^\p 2^\p}|} 
\Delta^{(3)}_{2^\p 2} .\\
%%%%%%%
\nonumber\\
%%%%%%%
&\frac{\dd \sigma_{1\times 2}^T}{\dd^2 \bp \, \dd^2 \bq\, \dd y_1 \, \dd y_2} = 
\frac{e^2 g^2 Q^2 N_c \sqrt{z_1z_2(1-z_1)(1-z_2)}}{(2\pi)^{10}}
\int\dd^{10} \bx 
\left\{\frac{N_c}{2} S_{12}S_{2^\p1^\p} +  C_F \left[1 - S_{12} - S_{2^\p1^\p}\right] - \frac{1}{2 N_c} S_{122^\p1^\p}\right\} 
\nonumber \\
&
\int\frac{\dd z}{z}  
e^{i\bp\cdot\bx_{1^\p 1}} e^{i\bq\cdot\bx_{2^\p 2}}
e^{i\frac{z}{z_1}\bp\cdot\bx_{31}}
e^{i\frac{z}{z_2}\bq\cdot\bx_{2^\p 3}} 
K_1(|\bx_{12}|Q_2)K_1(|\bx_{1^\p 2^\p}|Q_1) 
\nonumber \\
&
4\Re \Bigg[ \frac{(\bx_{12}\de)(\bx_{1^\p 2^\p}\des)}{|\bx_{12}||\bx_{1^\p2^\p}|}
\Bigg\{ (z_1^2+z_2^2)(1-z_1)(1-z_2) 
\frac{(\bx_{31}\de)(\bx_{2^\p 3}\des)}{\bx_{31}^2 \bx_{2^\p 3}^2} 
+ 
z_1z_2((1-z_1)^2+(1-z_2)^2)
\frac{(\bx_{31}\des)(\bx_{2^\p 3}\de)}{\bx_{31}^2 \bx_{2^\p 3}^2}
\Bigg\}\Bigg] .\\
%%%%%%%
\nonumber\\
%%%%%%%
&\frac{\dd \sigma_{3\times 3}^{T}}{\dd^2 \bp \, \dd^2 \bq\, \dd y_1 \, \dd y_2}  =
\frac{e^2 g^2 Q^2 N_c z_1 z_2^3}{(2\pi)^{10}} 
\int \dd^{10}\bx  
\left\{\frac{N_c}{2} \left[S_{11^\p} S_{22^\p} - S_{13} S_{32} - 
S_{31^\p} S_{2^\p3}\right] 
+ C_F - \frac{1}{2 N_c} \left[S_{122^\p1^\p} - S_{12} - S_{2^\p1^\p} \right] \right\} 
\nonumber \\
& 
\int\frac{\dd z}{z}  
e^{i\bp\cdot\bx_{1^\p 1}} e^{i\bq\cdot\bx_{2^\p 2}} 
\frac{K_1(QX)K_1(QX^\p)}{XX^\p} \, 4\Re \Bigg[(z_1^2+z_2^2)\frac{(\bx_{31}\de)(\bx_{3 1^\p}\des)}{\bx_{31}^2\bx_{3 1^\p}^2} [(z_1\bx_{12}+z\bx_{32})\de][(z_1\bx_{1^\p 2^\p}+z\bx_{3 2^\p})\des] \nonumber \\
&+\left((1-z_2)^2 + \frac{(z_1z_2)^2}{(1-z_2)^2}\right) \frac{(\bx_{31}\des)(\bx_{3 1^\p}\de)}{\bx_{31}^2\bx_{3 1^\p}^2}[(z_1\bx_{12}+z\bx_{32})\de][(z_1\bx_{1^\p 2^\p}+z\bx_{3 2^\p})\des] \nonumber \\
&-\frac{z_1^2 z_2 z}{2(1-z_2)^2}\left\{ \frac{(\bx_{31}\des)}{\bx_{31}^2}[(z_1\bx_{12}+z\bx_{32})\de] + \frac{(\bx_{31^\p}\de)}{\bx_{31^\p}^2}[(z_1\bx_{1^\p2^\p}+z\bx_{32^\p})\des]\right\} +\frac{z_1^2 z^2}{4(1-z_2)^2}\Bigg] .\\
%%%%%%%
\nonumber\\
%%%%%%%
&\frac{\dd \sigma_{4\times 4}^{T}}{\dd^2 \bp\, \dd^2 \bq\, \dd y_1 \, \dd y_2} =
\frac{e^2 g^2 Q^2 N_c z_2 z_1^3}{(2\pi)^{10}}
\int \dd^{10}\bx  
\left\{\frac{N_c}{2} \left[S_{11^\p} S_{22^\p} - S_{13} S_{32} - 
S_{31^\p} S_{2^\p3}\right] 
+ C_F - \frac{1}{2 N_c} \left[S_{122^\p1^\p} - S_{12} - S_{2^\p1^\p} \right] \right\} 
\nonumber \\
& 
\int\frac{\dd z}{z} 
e^{i\bp\cdot\bx_{1^\p 1}}e^{i\bq\cdot\bx_{2^\p 2}} 
\frac{K_1(QX)K_1(QX^\p)}{XX^\p}\, 4\Re \Bigg[(z_1^2+z_2^2)\frac{(\bx_{32}\de)(\bx_{3 2^\p}\des)}{\bx_{32}^2\bx_{3 2^\p}^2} [(z_2\bx_{21}+z\bx_{31})\de][(z_2\bx_{2^\p 1^\p}+z\bx_{3 1^\p})\des] \nonumber \\
&+\left((1-z_1)^2 + \frac{(z_1z_2)^2}{(1-z_1)^2}\right) \frac{(\bx_{32}\des)(\bx_{3 2^\p}\de)}{\bx_{32}^2\bx_{3 2^\p}^2}[(z_2\bx_{21}+z\bx_{31})\de][(z_2\bx_{2^\p 1^\p}+z\bx_{3 1^\p})\des] \nonumber \\
&-\frac{z_2^2 z_1 z}{2(1-z_1)^2}\left\{ \frac{(\bx_{32}\des)}{\bx_{32}^2}[(z_2\bx_{21}+z\bx_{31})\de] + \frac{(\bx_{32^\p}\de)}{\bx_{32^\p}^2}[(z_2\bx_{2^\p1^\p}+z\bx_{31^\p})\des]\right\} +\frac{z_2^2 z^2}{4(1-z_1)^2}\Bigg] .\\
%%%%%%%
\nonumber\\
%%%%%%%
&\frac{\dd \sigma_{3\times 4}^T}{\dd^2 \bp\,\dd^2\bq\,\dd y_1\, \dd y_2} = 
\frac{e^2 g^2 Q^2 N_c (z_1 z_2)^2}{(2\pi)^{10}} 
\int \dd^{10}\bx  
\left\{\frac{N_c}{2} \left[S_{11^\p} S_{22^\p} - S_{13} S_{32} - 
S_{31^\p} S_{2^\p3}\right] 
+ C_F - \frac{1}{2 N_c} \left[S_{122^\p1^\p} - S_{12} - S_{2^\p1^\p} \right] \right\} 
\nonumber \\
& 
\int\frac{\dd z}{z} 
e^{i\bp\cdot\bx_{1^\p 1}}e^{i\bq\cdot\bx_{2^\p 2}}
\frac{K_1(QX)K_1(QX^\p)}{XX^\p} \,
4\Re\Bigg[(z_1^2+z_2^2)\frac{(\bx_{31}\de)(\bx_{3 2^\p}\des)}{\bx_{31}^2\bx_{3 2^\p}^2} [(z_1\bx_{12}+z\bx_{32})\de][(z_2\bx_{2^\p 1^\p}+z\bx_{3 1^\p})\des] \nonumber \\
&+ \frac{z_1z_2}{(1-z_1)(1-z_2)}[(1-z_1)^2+(1-z_2)^2] \frac{(\bx_{31}\des)(\bx_{3 2^\p}\de)}{\bx_{31}^2\bx_{3 2^\p}^2}[(z_1\bx_{12}+z\bx_{32})\de][(z_2\bx_{2^\p 1^\p}+z\bx_{3 1^\p})\des] \nonumber \\
&-\frac{z_2 z(1-z_2)}{2(1-z_1)} \frac{(\bx_{31}\des)}{\bx_{31}^2}[(z_1\bx_{12}+z\bx_{32})\de]-\frac{z_1z(1-z_1)}{2(1-z_2)}\frac{(\bx_{3 2^\p}\de)}{\bx_{3 2^\p}^2} [(z_2\bx_{2^\p 1^\p}+z\bx_{3 1^\p})\des]\Bigg] .
\end{align}

\begin{align}
&\frac{\dd \sigma_{1\times 3}^T}{\dd^2 \bp\, \dd^2 \bq \, \dd y_1 \, \dd y_2} =  
\frac{-e^2 g^2 Q^2 N_c z_2^{5/2}\sqrt{1-z_2}}{(2\pi)^{10}} 
\int\dd^{10}\bx  
\left\{\frac{N_c}{2} \left[S_{122^\p3} S_{31^\p} - S_{31^\p} S_{2^\p3}\right] 
+ C_F \left[1 - S_{12}\right] - \frac{1}{2 N_c} \left[S_{122^\p1^\p} - S_{2^\p1^\p}\right]\right\}  
\nonumber \\
&
\int \frac{\dd z}{z} 
e^{i\bp\cdot\bx_{1^\p 1}} e^{i\bq\cdot\bx_{2^\p 2}} e^{i\frac{z}{z_1}\bp\cdot\bx_{31}} 
\frac{K_1(|\bx_{12}|Q_2) K_1(QX^\p)}{X^\p}4\Re \Bigg[ (1-z_2)(z_1^2+z_2^2)\frac{(\bx_{12}\de)(\bx_{3 1^\p}\des)}{|\bx_{12}|\bx_{3 1^\p}^2} \frac{(\bx_{31}\de)[(z_1\bx_{1^\p 2^\p}+z\bx_{3 2^\p})\des]}{\bx_{31}^2} \nonumber \\
&+\left((1-z_2)^3+\frac{(z_1z_2)^2}{1-z_2}\right) \frac{(\bx_{12}\de)(\bx_{3 1^\p}\de)}{|\bx_{12}|\bx_{3 1^\p}^2} \frac{(\bx_{31}\des)[(z_1\bx_{1^\p 2^\p}+z\bx_{3 2^\p})\des]}{\bx_{31}^2} -\frac{z_1^2 z_2 z}{2(1-z_2)} \frac{(\bx_{12}\de)}{|\bx_{12}|} \frac{(\bx_{31}\des)}{\bx_{31}^2}\Bigg] . 
\\
%%%%%%%
\nonumber\\
%%%%%%%
&\frac{\dd \sigma_{1\times 4}^T}{\dd^2 \bp\, \dd^2 \bq\, \dd y_1 \, \dd y_2} =  
\frac{-e^2 g^2 Q^2 N_c z_1 z_2^{3/2}\sqrt{1-z_2}}{(2\pi)^{10}} 
\int\dd^{10}\bx  
\left\{\frac{N_c}{2} \left[S_{122^\p3} S_{31^\p} - S_{31^\p} S_{2^\p3}\right] 
+ C_F \left[1 - S_{12}\right] - \frac{1}{2 N_c} \left[S_{122^\p1^\p} - S_{2^\p1^\p}\right]\right\}   
\nonumber \\
&
\int \frac{ \dd z}{z}
e^{i\bp\cdot\bx_{1^\p 1}} e^{i\bq\cdot\bx_{2^\p 2}} e^{i\frac{z}{z_1}\bp\cdot\bx_{31}} 
\frac{K_1(|\bx_{12}|Q_2) K_1(QX^\p)}{X^\p} 4\Re \Bigg[ (1-z_2)(z_1^2+z_2^2)\frac{(\bx_{12}\de)(\bx_{3 2^\p}\des)}{|\bx_{12}|\bx_{3 2^\p}^2} \frac{(\bx_{31}\de) [(z_2\bx_{2^\p 1^\p}+z \bx_{3 1^\p})\des]}{\bx_{31}^2} \nonumber \\
&+\frac{z_1z_2}{1-z_1}\left((1-z_1)^2+(1-z_2)^2\right)\frac{(\bx_{12}\de)(\bx_{3 2^\p}\de)}{|\bx_{12}|\bx_{3 2^\p}^2} \frac{(\bx_{31}\des) [(z_2\bx_{2^\p 1^\p}+z \bx_{3 1^\p})\des]}{\bx_{31}^2} -\frac{z_2 z(1-z_2)^2}{2(1-z_1)} \frac{(\bx_{12}\de)}{|\bx_{12}|} \frac{(\bx_{31}\des)}{\bx_{31}^2}\Bigg] . 
\\
%%%%%%%
\nonumber\\
%%%%%%%
&\frac{\dd \sigma_{2\times 3}^T}{\dd^2 \bp\, \dd^2\bq\, \dd y_1 \, \dd y_2} =  
\frac{e^2 g^2 Q^2 N_c z_2 z_1^{3/2}\sqrt{1-z_1}}{(2\pi)^{10}} 
\int\dd^{10}\bx 
\left\{\frac{N_c}{2} \left[S_{1231^\p} S_{2^\p3} - S_{31^\p} S_{2^\p3}\right] 
+ C_F \left[1 - S_{12}\right] - \frac{1}{2 N_c} \left[S_{122^\p1^\p} - S_{2^\p1^\p}\right]\right\}  
\nonumber \\
&
\int\frac{\dd z}{z} 
e^{i\bp\cdot\bx_{1^\p 1}} e^{i\bq\cdot\bx_{2^\p 2}} e^{i\frac{z}{z_2}\bq\cdot\bx_{32}}  
\frac{K_1(|\bx_{12}|Q_1) K_1(QX^\p)}{X^\p}4\Re \Bigg[ (1-z_1)(z_1^2+z_2^2)\frac{(\bx_{12}\de)(\bx_{3 1^\p}\des)}{|\bx_{12}|\bx_{3 1^\p}^2} \frac{(\bx_{32}\de) [(z_1\bx_{1^\p 2^\p}+z \bx_{3 2^\p})\des]}{\bx_{32}^2} \nonumber \\
&+\frac{z_1z_2}{1-z_2}\left((1-z_1)^2+(1-z_2)^2\right)\frac{(\bx_{12}\de)(\bx_{3 1^\p}\de)}{|\bx_{12}|\bx_{3 1^\p}^2} \frac{(\bx_{32}\des) [(z_1\bx_{1^\p 2^\p}+z \bx_{3 2^\p})\des]}{\bx_{32}^2} -\frac{z_1 z(1-z_1)^2}{2(1-z_2)} \frac{(\bx_{12}\de)}{|\bx_{12}|} \frac{(\bx_{32}\des)}{\bx_{32}^2}\Bigg] . 
\\
%%%%%%%
\nonumber\\
%%%%%%%
&\frac{\dd \sigma_{2\times 4}^T}{\dd^2 \bp\, \dd^2 \bq\, \dd y_1 \, \dd y_2} =  
\frac{e^2 g^2 Q^2 N_c z_1^{5/2}\sqrt{1-z_1}}{(2\pi)^{10}} 
\int\dd^{10}\bx 
\left\{\frac{N_c^2}{2} \left[S_{1231^\p} S_{2^\p3} - S_{31^\p} S_{2^\p3}\right] 
+ C_F \left[1 - S_{12}\right] - \frac{1}{2 N_c} \left[S_{122^\p1^\p} - S_{2^\p1^\p}\right]\right\}    
\nonumber \\
&
\int \frac{\dd z}{z}
e^{i\bp\cdot\bx_{1^\p 1}} e^{i\bq\cdot\bx_{2^\p 2}} e^{i\frac{z}{z_2}\bq\cdot\bx_{32}} 
\frac{K_1(|\bx_{12}|Q_1) K_1(QX^\p)}{X^\p}4\Re \Bigg[ (1-z_1)(z_1^2+z_2^2)\frac{(\bx_{12}\de)(\bx_{3 2^\p}\des)}{|\bx_{12}|\bx_{3 2^\p}^2} \frac{(\bx_{32}\de)[(z_2\bx_{2^\p 1^\p}+z\bx_{3 1^\p})\des]}{\bx_{32}^2} \nonumber \\
&+\left((1-z_1)^3+\frac{(z_1z_2)^2}{1-z_1}\right) \frac{(\bx_{12}\de)(\bx_{3 2^\p}\de)}{|\bx_{12}|\bx_{3 2^\p}^2} \frac{(\bx_{32}\des)[(z_2\bx_{2^\p 1^\p}+z\bx_{3 1^\p})\des]}{\bx_{32}^2} -\frac{z_2^2 z_1 z}{2(1-z_1)} \frac{(\bx_{12}\de)}{|\bx_{12}|} \frac{(\bx_{32}\des)}{\bx_{32}^2}\Bigg] . 
\end{align}
}

\begin{align}
&\frac{\dd \sigma^{T}_5}{\dd^2 \bp \, \dd^2 \bq \, \dd y_1 \, \dd y_2} =  
\frac{e^2 g^2 Q^2 N_c z_2^{5/2} \sqrt{z_1}}{(2\pi)^{10} }  
\int\dd^{10} \bx \,
\left\{\frac{N_c}{2} \left[S_{13}  S_{322^\p1^\p} - S_{32} S_{13}\right] 
+ C_F \left[1 - S_{2^\p1^\p}\right] - 
\frac{1}{2 N_c} \left[S_{122^\p1^\p} - S_{12}\right]\right\}
\nonumber \\
&
\int_0^{z_1} \frac{\dd z}{z} 
e^{i\bp\cdot\bx_{1^\p 1}} e^{i\bq\cdot\bx_{2^\p 2}} 
e^{-i\frac{z}{z_1} \bp\cdot\bx_{3 1}} 
\frac{K_1(QX_5)K_1(|\bx_{1^\p 2^\p}|Q_1)}{X_5\bx_{31}^2 |\bx_{1^\p 2^\p}|} 
\nonumber \\
& \bx_{1^\p 2^\p}\cdot\left[(z_1^2+z_2^2)(z_1^2+(z_1-z)^2)\bx_{32} +(z_1-z)
\left(z_1(z_1^2+(z_1-z)^2)+z_2[z_1z_2+(z_1-z)(z_2+z)]\right)\bx_{13}\right] .
\end{align}

\begin{align}
&\frac{\dd \sigma^{T}_6}{\dd^2 \bp \, \dd^2 \bq \, \dd y_1 \, \dd y_2} = 
-\frac{e^2 g^2 Q^2 N_c z_1^{5/2} \sqrt{z_2}}{(2\pi)^{10}}  
\int\dd^{10} \bx 
\left\{\frac{N_c}{2} \left[S_{2^\p1^\p13} S_{32} - S_{13} S_{32}\right] 
+ C_F \left[1 - S_{2^\p1^\p}\right] - 
\frac{1}{2 N_c} \left[S_{122^\p1^\p} - S_{12}\right]\right\}
\nonumber \\
&
\int_0^{z_2} \frac{\dd z}{z} 
e^{i\bp\cdot\bx_{1^\p 1}} e^{i\bq\cdot\bx_{2^\p 2}}
e^{-i\frac{z}{z_2}\bq\cdot\bx_{3 2}}
\frac{K_1(QX_6) K_1(|\bx_{1^\p 2^\p}| Q_1)}{X_6 |\bx_{1^\p 2^\p}|\bx_{32}^2} 
\nonumber \\
&
\bx_{1^\p 2^\p}\cdot\left[(z_1^2+z_2^2)(z_2^2+(z_2-z)^2)\bx_{31}+(z_2-z)[z_2(z_2^2+(z_2-z)^2)+z_1(z_1z_2+(z_2-z)(z_1+z))]\bx_{23}\right].
\end{align}

\begin{align}
&\frac{\dd \sigma^{T}_{7}}{\dd^2 \bp \, \dd^2 \bq \, \dd y_1 \, \dd y_2} = 
\frac{e^2 g^2 Q^2 N_c (z_1z_2)^{3/2}}{(2\pi)^{10}}  
\int \dd^{10} \bx\,  
\left\{\frac{N_c}{2} \left[S_{13} S_{322^\p1^\p} - S_{32} S_{13}\right] 
+ C_F \left[1 - S_{2^\p1^\p}\right] - 
\frac{1}{2 N_c} \left[S_{122^\p1^\p} - S_{12}\right]\right\} 
\nonumber \\
&
\int_0^{z_1}\frac{\dd z \, (z_1-z)}{z}
e^{i\bp\cdot\bx_{1^\p 1}} e^{i\bq\cdot\bx_{2^\p 2}} 
e^{-i\frac{z}{z_1}\bp\cdot\bx_{3 1}} 
\frac{ K_1(QX_5)  K_1(|\bx_{1^\p2^\p}|Q_1)}{X_5 \bx_{31}^2 |\bx_{1^\p 2^\p}|}
\nonumber \\
&
\Bigg[ \frac{4\Re}{\bx_{32}^2}\Bigg\{ (\bx_{1^\p 2^\p}\des)\left[\left(\bx_{31}+\frac{z_2}{z_2+z}\bx_{23}\right)\de\right]\Bigg[\frac{z_2 (z_1-z)}{z_1}\left(z_1^2+(z_2+z)^2\right)(\bx_{31}\de)(\bx_{32}\des) + 
\nonumber \\
& 
(z_2+z)(z_2^2+(z_1-z)^2)(\bx_{32}\de)(\bx_{31}\des)\Bigg]\Bigg\} 
-\frac{z_1 z_2z}{z_2+z}\bx_{31}\cdot\bx_{1^\p 2^\p}\Bigg].
\end{align}

\begin{align}
&\frac{\dd \sigma^{T}_{8}}{\dd^2 \bp \, \dd^2 \bq \, \dd y_1 \, \dd y_2} =  
\frac{- e^2 g^2 Q^2 N_c (z_1 z_2)^{3/2}}{(2\pi)^{10}}  
\dd^{10}\bx\,  
\left\{\frac{N_c}{2} \left[S_{2^\p1^\p13} S_{32} - S_{13} S_{32}\right] 
+ C_F \left[1 - S_{2^\p1^\p}\right] - 
\frac{1}{2 N_c} \left[S_{122^\p1^\p} - S_{12}\right]\right\}
\nonumber \\
&
\int_0^{z_2} \frac{\dd z \, (z_2-z)}{z} 
e^{i\bp\cdot\bx_{1^\p 1}} e^{i\bq\cdot\bx_{2^\p 2}} 
e^{-i\frac{z}{z_2}\bq\cdot\bx_{3 2}} 
\frac{K_1(QX_6)K_1(|\bx_{1^\p2^\p}|Q_1)}{X_6 |\bx_{1^\p 2^\p}| \bx_{32}^2} 
\nonumber \\
&
\Bigg[ \frac{4\Re}{\bx_{31}^2} \Bigg\{(\bx_{1^\p 2^\p}\des)\left[\left(\bx_{32}+\frac{z_1}{z_1+z}\bx_{13}\right)\de\right]
\Bigg[ (z_1+z)(z_1^2+(z_2-z)^2)(\bx_{31}\de)(\bx_{32}\des) + 
\nonumber \\
& 
\frac{z_1 (z_2-z)}{z_2}\left(z_2^2+(z_1+z)^2\right)(\bx_{32}\de)(\bx_{31}\des)\Bigg]\Bigg\}-\frac{z_1z_2z}{z_1+z}\bx_{32}\cdot\bx_{1^\p 2^\p}\Bigg].
\end{align}

\begin{align}
\frac{\dd \sigma_9^{T}}{\dd^2 \bp \, \dd^2 \bq\, \dd y_1\, \dd y_2} = &
\frac{-e^2 g^2 Q^2 N_c (z_1 z_2)^2(z_1^2+z_2^2)}{2 (2\pi)^8} 
\int \dd^8 \bx \, C_F \left\{ S_{122^\p1^\p} - S_{12} - S_{2^\p1^\p} + 1\right\}  
\frac{\bx_{12}\cdot\bx_{1^\p 2^\p}}{|\bx_{12}||\bx_{1^\p 2^\p}|}  
\nonumber \\
&
e^{i\bp\cdot\bx_{1^\p 1}} e^{i\bq\cdot\bx_{2^\p 2}}  
K_1(|\bx_{12}|Q_1) K_1(|\bx_{1^\p 2^\p}|Q_1) 
\int_0^{z_1} \frac{\dd z}{z}\left[ \frac{z_1^2+(z_1-z)^2}{z_1^2}\right] \int \dtwo{\bk} \frac{1}{\left(\bk-\frac{z}{z_1}\bp\right)^2} .
\end{align}

\begin{align}
\frac{\dd \sigma_{10}^{T}}{\dd^2 \bp \, \dd^2 \bq\, \dd y_1\, \dd y_2} = &
\frac{-e^2 g^2 Q^2 N_c (z_1 z_2)^2(z_1^2+z_2^2)}{2 (2\pi)^8} 
\int \dd^8 \bx \, C_F \left\{ S_{122^\p1^\p} - S_{12} - S_{2^\p1^\p} + 1\right\} 
\frac{\bx_{12}\cdot\bx_{1^\p 2^\p}}{|\bx_{12}||\bx_{1^\p 2^\p}|}  
\nonumber \\
& 
e^{i\bp\cdot\bx_{1^\p 1}} e^{i\bq\cdot\bx_{2^\p 2}}
K_1(|\bx_{12}|Q_1) K_1(|\bx_{1^\p 2^\p}|Q_1)
\int_0^{z_2} \frac{\dd z}{z}
\left[ \frac{z_2^2+(z_2-z)^2}{z_2^2}\right] \int \dtwo{\bk} 
\frac{1}{\left(\bk-\frac{z}{z_2}\bq\right)^2} .
\end{align}

\begin{align}
&\frac{\dd \sigma_{11}^{T}}{\dd^2 \bp\, \dd^2 \bq\, \dd y_1 \, \dd y_2} = 
\frac{i e^2 g^2 Q N_c z_2^{3/2}\sqrt{z_1}(z_1^2+z_2^2)}{(2\pi)^7} 
\int\dd^8 \bx \, C_F \left\{ S_{122^\p1^\p} - S_{12} - S_{2^\p1^\p} + 1\right\} 
\nonumber\\
&
e^{i\bp\cdot\bx_{1^\p 1}} e^{i\bq\cdot\bx_{2^\p 2}}
\frac{K_1(|\bx_{1^\p 2^\p}|Q_1)}{|\bx_{1^\p 2^\p}|}  
\int_0^{z_1} \frac{\dd z}{z^2} \frac{[(z_1-z)^2+z_1^2]}{(z_1-z)} 
\int \dtwo{\bk_2} \int \dtwo{\bk_1}  \frac{\bk_1\cdot\bx_{1^\p 2^\p}}
{\big[\bk_1^2+Q_1^2\big]
\left[Q^2 +\frac{\bk_1^2}{z_1z_2} +\frac{z_1}{z(z_1-z)}\bk_2^2\right]}
e^{i\bk_1\cdot\bx_{1 2}} . 
\end{align}

\begin{align}
&\frac{\dd \sigma_{12}^{T}}{\dd^2 \bp\, \dd^2 \bq\, \dd y_1 \, \dd y_2} = 
\frac{-i e^2 g^2 Q N_c z_1^{3/2}\sqrt{z_2}(z_1^2+z_2^2)}{(2\pi)^7} 
\int\dd^8 \bx \, C_F \left\{ S_{122^\p1^\p} - S_{12} - S_{2^\p1^\p} + 1\right\}    
\nonumber\\
&
e^{i\bp\cdot\bx_{1^\p 1}} e^{i\bq\cdot\bx_{2^\p 2}} 
\frac{K_1(|\bx_{1^\p 2^\p}|Q_1)}{|\bx_{1^\p 2^\p}|} 
\int_0^{z_2} \frac{\dd z}{z^2} 
\frac{[(z_2-z)^2+z_2^2]}{(z_2-z)} \int \dtwo{\bk_2} 
\int \dtwo{\bk_1}  \frac{\bk_1\cdot\bx_{1^\p 2^\p}}
{\big[\bk_1^2+Q_1^2\big]\left[Q^2 +\frac{\bk_1^2}{z_1z_2} +\frac{z_2}{z(z_2-z)}\bk_2^2\right]} 
e^{i\bk_1\cdot\bx_{2 1}} .
\end{align}

\begin{align}
&\frac{\dd \sigma_{13(1)}^{T}}{\dd^2 \bp\, \dd^2 \bq\, \dd y_1\, \dd y_2} =  \frac{e^2 g^2 Q^2 N_c(z_1z_2)^{3/2}}{(2\pi)^8} 
\int \dd^8\bx\,
\left\{\frac{N_c}{2} \left[S_{2^\p1^\p} S_{12} \right] 
+ C_F \left[1 - S_{2^\p1^\p} - S_{12}\right] - 
\frac{1}{2 N_c} \left[S_{122^\p1^\p}\right]\right\}  
\nonumber \\
& 
\int_{0}^{z_2} \dd z \sqrt{(z_1+z)(z_2-z)}  \, 
e^{i\bp\cdot\bx_{1^\p 1}} e^{i\bq\cdot\bx_{2^\p 2}} 
\frac{ K_1\left(|\bx_{12}|Q\sqrt{(z_1+z)(z_2-z)}\right)}{|\bx_{12}|} \frac{K_1(|\bx_{1^\p 2^\p}|Q_1)}{|\bx_{1^\p 2^\p}|} \int \dtwo{\bk}e^{i\bk\cdot\bx_{21}}
\nonumber \\
&
4\Re\Bigg[ (\bx_{12}\de)(\bx_{1^\p 2^\p}\des)\Bigg\{ \frac{\frac{z_2(z_2-z)[z_1(z_1+z)+z_2(z_2-z)]}{2z}}{(z_2\bk-z\bq)^2} + 
\nonumber \\
&
z_1z_2 z\left(\frac{ z_1z\left(\frac{\bp\de}{z_1}-\frac{\bq\de}{z_2}\right)\left(\frac{\bp\des}{z_1}-\frac{\bk\des}{z}\right) +z^2\left(\frac{\bk\de}{z}-\frac{\bq\de}{z_2}\right)\left(\frac{\bp\des}{z_1}-\frac{\bk\des}{z}\right) -z_2z\left(\frac{\bk\de}{z}-\frac{\bq\de}{z_2}\right)\left(\frac{\bp\des}{z_1}-\frac{\bq\des}{z_2}\right) - p\cdot q}{(z_2\bk-z\bq)^2\left[\frac{(z_1\bk-z\bp)^2}{z_1(z_1+z)}-\frac{(z_2\bk-z\bq)^2}{z_2 (z_2-z)}\right]}\right)  +
\nonumber \\
&
z_2^2 z (z_2-z)\left(\frac{ z_1z\left(\frac{\bp\des}{z_1}-\frac{\bq\des}{z_2}\right)\left(\frac{\bp\de}{z_1}-\frac{\bk\de}{z}\right) +z^2\left(\frac{\bk\des}{z}-\frac{\bq\des}{z_2}\right)\left(\frac{\bp\de}{z_1}-\frac{\bk\de}{z}\right) -z_2z\left(\frac{\bk\des}{z}-\frac{\bq\des}{z_2}\right)\left(\frac{\bp\de}{z_1}-\frac{\bq\de}{z_2}\right) - p\cdot q}{(z_1+z)(z_2\bk-z\bq)^2\left[\frac{(z_1\bk-z\bp)^2}{z_1(z_1+z)}-\frac{(z_2\bk-z\bq)^2}{z_2 (z_2-z)}\right]}\right)\Bigg\}\Bigg]. 
\end{align}

\begin{align}
&\frac{\dd \sigma_{13(2)}^{T}}{\dd^2 \bp\, \dd^2 \bq\, \dd y_1\, \dd y_2} =  \frac{e^2 g^2 Q^2 N_c (z_1z_2)^{3/2}}{(2\pi)^8}
\int \dd^8\bx\,
\left\{\frac{N_c}{2} \left[S_{2^\p1^\p} S_{12} \right] 
+ C_F \left[1 - S_{2^\p1^\p} - S_{12}\right] - 
\frac{1}{2 N_c} \left[S_{122^\p1^\p}\right]\right\}  
\nonumber \\
& 
\int_{0}^{z_1} \dd z \sqrt{(z_1-z)(z_2+z)} \, 
e^{i\bp\cdot\bx_{1^\p 1}}e^{i\bq\cdot\bx_{2^\p 2}}
\frac{ K_1\left(|\bx_{12}|Q\sqrt{(z_1-z)(z_2+z)}\right)}{|\bx_{12}|} 
\frac{K_1(|\bx_{1^\p 2^\p}|Q_1)}{|\bx_{1^\p 2^\p}|} 
\int \dtwo{\bk} e^{i\bk\cdot\bx_{12}} 
\nonumber \\
&
4\Re\Bigg[  (\bx_{12}\de)(\bx_{1^\p 2^\p}\des) 
\Bigg\{ \frac{ \frac{z_1(z_1-z)[z_1(z_1-z)+z_2(z_2+z)]}{2z}}{(z_1\bk-z\bp)^2}  +     
\nonumber \\
&
z_1z_2z\left(\frac{ z_1z\left(\frac{\bp\des}{z_1}-\frac{\bq\des}{z_2}\right)\left(\frac{\bp\de}{z_1}-\frac{\bk\de}{z}\right) -z^2\left(\frac{\bk\des}{z}-\frac{\bq\des}{z_2}\right)\left(\frac{\bp\de}{z_1}-\frac{\bk\de}{z}\right) -z_2z\left(\frac{\bk\des}{z}-\frac{\bq\des}{z_2}\right)\left(\frac{\bp\de}{z_1}-\frac{\bq\de}{z_2}\right) + p\cdot q}{(z_1\bk-z\bp)^2\left[\frac{(z_1\bk-z\bp)^2}{z_1(z_1-z)}-\frac{(z_2\bk-z\bq)^2}{z_2 (z_2+z)}\right]}\right) + 
\nonumber \\
&
z_1^2 z(z_1-z)\left(\frac{ z_1z\left(\frac{\bp\de}{z_1}-\frac{\bq\de}{z_2}\right)\left(\frac{\bp\des}{z_1}-\frac{\bk\des}{z}\right) -z^2\left(\frac{\bk\de}{z}-\frac{\bq\de}{z_2}\right)\left(\frac{\bp\des}{z_1}-\frac{\bk\des}{z}\right) -z_2z\left(\frac{\bk\de}{z}-\frac{\bq\de}{z_2}\right)\left(\frac{\bp\des}{z_1}-\frac{\bq\des}{z_2}\right) + p\cdot q}{(z_2+z)(z_1\bk-z\bp)^2\left[\frac{(z_1\bk-z\bp)^2}{z_1(z_1-z)}-\frac{(z_2\bk-z\bq)^2}{z_2 (z_2+z)}\right]}\right)\Bigg\}\Bigg].
\end{align}
\eq{
&\frac{ \dd \sigma_{14(1)}^{T}}{\dd^2 \bp \, \dd^2 \bq \, \dd y_1 \, \dd y_2} = 
\frac{-i e^2 g^2 Q N_c (z_1z_2)^{3/2}}{(2\pi)^7} 
\int\dd^8 \bx \, C_F \left\{ S_{122^\p1^\p} - S_{12} - S_{2^\p1^\p} + 1\right\} 
\frac{ K_1(|\bx_{1^\p 2^\p}|Q_1)}{|\bx_{1^\p 2^\p}|}
e^{i\bp\cdot\bx_{1^\p 1}} e^{i \bq\cdot\bx_{2^\p 2}} 
\nonumber \\
&
\int_0^{z_1} \frac{\dd z}{z} 
\int\dtwo{\bk_1}\dtwo{\bk_2}e^{i\bk_2\cdot\bx_{12}} \Bigg\{\frac{\br{z_1(z_1-z)+z_2(z_2+z)-z(1-z)} (\bk_2\cdot\bx_{1^\p 2^\p})    }
{\Big[\bk_2^2 + Q_1^2\Big] \left[\pa{\bk_1-\frac{z_1-z}{z_1}\bk_2}^2+\frac{z(z_1-z)}{z_2 z_1^2} \bk_2^2 + \frac{z}{z_1}(z_1-z)Q^2\right] } + 
\nonumber \\
&
\frac{\frac{(z_1-z)}{z_1}(1+z^2-2z_2(z_1-z))(\bk_1\cdot\bx_{1^\p 2^\p}) }{\Big[\bk_1^2 + (z_1-z)(z_2+z)Q^2\Big] \left[\left(\bk_1-\frac{z_1-z}{z_1}\bk_2\right)^2+\frac{z(z_1-z)}{z_2 z_1^2} \bk_2^2 + \frac{z}{z_1}(z_1-z)Q^2\right]} - 
\nonumber \\
&
\frac{Q^2 \frac{(z_1-z)}{z_1}\br{2z_1z_2z(\bk_1\cdot\bx_{1^\p 2^\p}) + z(z+z_2-z_1)^2(\bk_2\cdot\bx_{1^\p 2^\p})}}{\Big[\bk_1^2 + (z_1-z)(z_2+z)Q^2\Big]\Big[\bk_2^2 + Q_1^2\Big]\left[\left(\bk_1-\frac{z_1-z}{z_1}\bk_2\right)^2+\frac{z(z_1-z)}{z_2 z_1^2} \bk_2^2 + \frac{z}{z_1}(z_1-z)Q^2\right]}
\Bigg\} . 
}

\eq{
&\frac{ \dd \sigma_{14(2)}^{T}}{\dd^2 \bp \, \dd^2 \bq \, \dd y_1 \, \dd y_2} = 
\frac{-i e^2 g^2 Q N_c (z_1z_2)^{3/2}}{(2\pi)^7} 
\int\dd^8 \bx
C_F \left\{ S_{122^\p1^\p} - S_{12} - S_{2^\p1^\p} + 1\right\} 
\frac{ K_1(|\bx_{1^\p 2^\p}|Q_1)}{|\bx_{1^\p 2^\p}|}
e^{i\bp\cdot\bx_{1^\p 1}} e^{i\bq\cdot\bx_{2^\p 2}} 
\nonumber \\
&
\int_0^{z_2} \frac{\dd z}{z} 
\int  \dtwo{\bk_1}\dtwo{\bk_2}e^{i\bk_2\cdot\bx_{12}} 
\Bigg\{\frac{\br{z_2(z_2-z)+z_1(z_1+z)-z(1-z)} (\bk_2\cdot\bx_{1^\p 2^\p})    }
{\Big[\bk_2^2 + Q_1^2\Big] \left[\pa{\bk_1-\frac{z_2-z}{z_2}\bk_2}^2 +
\frac{z(z_2-z)}{z_1 z_2^2} \bk_2^2 + \frac{z}{z_2}(z_2-z)Q^2\right] } + 
\nonumber \\
&
\frac{\frac{(z_2-z)}{z_2}(1+z^2-2z_1(z_2-z))(\bk_1\cdot\bx_{1^\p 2^\p}) }
{\Big[\bk_1^2 + (z_2-z)(z_1+z)Q^2\Big] 
\left[\left(\bk_1-\frac{z_2-z}{z_2}\bk_2\right)^2+\frac{z(z_2-z)}{z_1 z_2^2} 
\bk_2^2 + \frac{z}{z_2}(z_2-z)Q^2\right]}  - 
\nonumber \\
& 
\frac{Q^2 \frac{(z_2-z)}{z_2}\br{2z_1z_2z(\bk_1\cdot\bx_{1^\p 2^\p}) + 
z(z+z_1-z_2)^2(\bk_2\cdot\bx_{1^\p 2^\p})}}{\Big[\bk_1^2 + (z_2-z)(z_1+z)Q^2\Big]\Big[\bk_2^2 + Q_1^2\Big]
\left[\left(\bk_1-\frac{z_2-z}{z_2}\bk_2\right)^2+\frac{z(z_2-z)}{z_1 z_2^2} 
\bk_2^2 + \frac{z}{z_2}(z_2-z)Q^2\right]}
\Bigg\}
}

\section{$z\rightarrow0$ limit of the NLO SIDIS cross-section from the transverse photon}
\label{App:z_to_zero}
%{\bf this section is corrected by Yu}

%{\bf confirmed jjm}

\subsubsection{real corrections}

\begin{align}
 & \frac{\dd\sigma_{1\times1}^{T}}{\dd^{2}\bp\,\dd y_{1}}=2\frac{e^{2}g^{2}Q^{2}N_{c}}{(2\pi)^{8}}\,z_{1}^{2}z_{2}\left(z_{1}^{2}+z_{2}^{2}\right)\int\frac{\dd z}{z}\int\dd^{8}\bx\frac{\bx_{12}\cdot\bx_{1^{\p}2}}{|\bx_{12}||\bx_{1^{\p}2}|}K_{1}(|\bx_{12}|Q_{1})K_{1}(|\bx_{1^{\p}2}|Q_{1})e^{i\bp\cdot\bx_{1^{\p}1}}\nonumber \\
 & C_{F}[S_{11^{\p}}-S_{12}-S_{21^{\p}}+1]\left(\frac{1}{2}\frac{1}{\bx_{13}^{2}}+\frac{1}{2}\frac{1}{\bx_{1^{\p}3}^{2}}-\frac{1}{2}\frac{\bx_{11^{\p}}^{2}}{\bx_{13}^{2}\bx_{1^{\p}3}^{2}}\right).
\end{align}
%%%%%%%%%%%%%%%%%%%%
%\newpage
%%%%%%%%%%%%%%%%%%%%
\begin{align}
 & \frac{\dd\sigma_{1\times2}^{T}}{\dd^{2}\bp\,\dd y_{1}}=-2\frac{e^{2}g^{2}Q^{2}N_{c}}{(2\pi)^{8}}z_{1}^{2}z_{2}(z_{1}^{2}+z_{2}^{2})\int\frac{\dd z}{z}\int\dd^{8}\bx\frac{\bx_{12}\cdot\bx_{1^{\p}2}}{|\bx_{12}||\bx_{1^{\p}2}|}K_{1}(|\bx_{12}|Q_{1})K_{1}(|\bx_{1^{\p}2}|Q_{1})e^{i\bp\cdot\bx_{1^{\p}1}}\nonumber \\
 & \bigg[C_{F}[S_{12}S_{21^{\p}}-S_{12}-S_{21^{\p}}+1]-\frac{1}{2N_{c}}\left(S_{11^{\p}}-S_{12}S_{21^{\p}}\right)\bigg]\left(\frac{1}{2}\frac{1}{\bx_{13}^{2}}+\frac{1}{2}\frac{1}{\bx_{23}^{2}}-\frac{1}{2}\frac{\bx_{12}^{2}}{\bx_{13}^{2}\bx_{23}^{2}}\right).
\end{align}

\begin{align}
\frac{\dd\sigma_{2\times1}^{T}}{\dd^{2}\bp\,\dd y_{1}} & =-2\frac{e^{2}g^{2}Q^{2}N_{c}}{(2\pi)^{8}}z_{1}^{2}z_{2}(z_{1}^{2}+z_{2}^{2})\int\frac{\dd z}{z}\int\dd^{8}\bx\frac{\bx_{12}\cdot\bx_{1^{\p}2}}{|\bx_{12}||\bx_{1^{\p}2}|}K_{1}(|\bx_{12}|Q_{1})K_{1}(|\bx_{1^{\p}2}|Q_{1})e^{i\bp\cdot\bx_{11^{\p}}}\nonumber \\
 & \bigg[C_{F}[S_{21}S_{1^{\p}2}-S_{21}-S_{1^{\p}2}+1]-\frac{1}{2N_{c}}\left(S_{1^{\p}1}-S_{21}S_{1^{\p}2}\right)\bigg]\left(\frac{1}{2}\frac{1}{\bx_{13}^{2}}+\frac{1}{2}\frac{1}{\bx_{23}^{2}}-\frac{1}{2}\frac{\bx_{12}^{2}}{\bx_{13}^{2}\bx_{23}^{2}}\right).
\end{align}

%%%%%%%%%%%%%%%%%%%%
%\newpage
%%%%%%%%%%%%%%%%%%%%
\begin{align}
 & \frac{\dd\sigma_{3\times3}^{T}}{\dd^{2}\bp\,\dd y_{1}}=2\frac{e^{2}g^{2}Q^{2}N_{c}}{(2\pi)^{8}}z_{1}^{2}z_{2}\left(z_{1}^{2}+z_{2}^{2}\right)\int\frac{\dd z}{z}\int\dd^{8}\bx\frac{\bx_{12}\cdot\bx_{1^{\p}2}}{|\bx_{12}||\bx_{1^{\p}2}|}K_{1}(|\bx_{12}|Q_{1})K_{1}(|\bx_{1^{\p}2}|Q_{1})\,e^{i\bp\cdot\bx_{1^{\p}1}}\nonumber \\
 & \bigg\{ C_{F}[S_{11^{\p}}-S_{13}S_{32}-S_{31^{\p}}S_{23}+1]-\frac{1}{2N_{c}}[S_{13}S_{32}+S_{31^{\p}}S_{23}-S_{12}-S_{21^{\p}}]\bigg\}\left(\frac{1}{2}\frac{1}{\bx_{13}^{2}}+\frac{1}{2}\frac{1}{\bx_{1^{\p}3}^{2}}-\frac{1}{2}\frac{\bx_{11^{\p}}^{2}}{\bx_{13}^{2}\bx_{1^{\p}3}^{2}}\right).
\end{align}
%%%%%%%%%%%%%%%%%%%%
%\newpage
%%%%%%%%%%%%%%%%%%%%
\begin{align}
 & \frac{\dd\sigma_{4\times4}^{T}}{\dd^{2}\bp\,\dd y_{1}}=2\frac{e^{2}g^{2}Q^{2}N_{c}}{(2\pi)^{8}}z_{1}^{2}z_{2}\left(z_{1}^{2}+z_{2}^{2}\right)\int\frac{\dd z}{z}\int\dd^{8}\bx\frac{\bx_{12}\cdot\bx_{1^{\p}2}}{|\bx_{12}||\bx_{1^{\p}2}|}K_{1}(|\bx_{12}|Q_{1})K_{1}(|\bx_{1^{\p}2}|Q_{1})e^{i\bp\cdot\bx_{1^{\p}1}}\nonumber \\
 & \bigg\{ C_{F}[S_{11^{\p}}-S_{13}S_{32}-S_{31^{\p}}S_{23}+1]-\frac{1}{2N_{c}}[S_{13}S_{32}+S_{31^{\p}}S_{23}-S_{12}-S_{21^{\p}}]\bigg\}\frac{1}{\bx_{23}^{2}}.
\end{align}

%%%%%%%%%%%%%%%%%%%%
%\newpage
%%%%%%%%%%%%%%%%%%%%
\begin{align}
 & \frac{\dd\sigma_{3\times4}^{T}}{\dd^{2}\bp\,\dd y_{1}}=-2\frac{e^{2}g^{2}Q^{2}N_{c}}{(2\pi)^{8}}z_{1}^{2}z_{2}(z_{1}^{2}+z_{2}^{2})\int\frac{\dd z}{z}\int\dd^{8}\bx\frac{\bx_{12}\cdot\bx_{1^{\p}2}}{|\bx_{12}||\bx_{1^{\p}2}|}K_{1}(|\bx_{12}|Q_{1})K_{1}(|\bx_{1^{\p}2}|Q_{1})e^{i\bp\cdot\bx_{1^{\p}1}}\nonumber \\
 & \bigg\{ C_{F}[S_{11^{\p}}-S_{13}S_{32}-S_{31^{\p}}S_{23}+1]-\frac{1}{2N_{c}}[S_{13}S_{32}+S_{31^{\p}}S_{23}-S_{12}-S_{21^{\p}}]\bigg\}\left(\frac{1}{2}\frac{1}{\bx_{13}^{2}}+\frac{1}{2}\frac{1}{\bx_{23}^{2}}-\frac{1}{2}\frac{\bx_{12}^{2}}{\bx_{13}^{2}\bx_{23}^{2}}\right).
\end{align}

%%%%%%%%%%%%%%%%%%%%
%%%%%%%%%%%%%%%%%%%%
\begin{align}
 & \frac{\dd\sigma_{4\times3}^{T}}{\dd^{2}\bp\,\dd y_{1}}=-2\frac{e^{2}g^{2}Q^{2}N_{c}}{(2\pi)^{8}}z_{1}^{2}z_{2}(z_{1}^{2}+z_{2}^{2})\int\frac{\dd z}{z}\int\dd^{8}\bx\frac{\bx_{12}\cdot\bx_{1^{\p}2}}{|\bx_{12}||\bx_{1^{\p}2}|}K_{1}(|\bx_{12}|Q_{1})K_{1}(|\bx_{1^{\p}2}|Q_{1})e^{i\bp\cdot\bx_{11^{\p}}}\nonumber \\
 & \bigg\{ C_{F}[S_{1^{\p}1}-S_{31}S_{23}-S_{1^{\p}3}S_{32}+1]-\frac{1}{2N_{c}}[S_{31}S_{23}+S_{1^{\p}3}S_{32}-S_{21}-S_{1^{\p}2}]\bigg\}\left(\frac{1}{2}\frac{1}{\bx_{13}^{2}}+\frac{1}{2}\frac{1}{\bx_{23}^{2}}-\frac{1}{2}\frac{\bx_{12}^{2}}{\bx_{13}^{2}\bx_{23}^{2}}\right).
\end{align}

%%%%%%%%%%%%%%%%%%%%
%\newpage
%%%%%%%%%%%%%%%%%%%%
\begin{align}
 & \frac{\dd\sigma_{1\times3}^{T}}{\dd^{2}\bp\,\dd y_{1}}=-2\frac{e^{2}g^{2}Q^{2}N_{c}}{(2\pi)^{8}}z_{1}^{2}z_{2}(z_{1}^{2}+z_{2}^{2})\int\frac{\dd z}{z}\int\dd^{8}\bx\frac{\bx_{12}\cdot\bx_{1^{\p}2}}{|\bx_{12}||\bx_{1^{\p}2}|}K_{1}(|\bx_{12}|Q_{1})K_{1}(|\bx_{1^{\p}2}|Q_{1})e^{i\bp\cdot\bx_{1^{\p}1}}\nonumber \\
 & \bigg\{ C_{F}[S_{13}S_{31^{\p}}-S_{31^{\p}}S_{23}-S_{12}+1]-\frac{1}{2N_{c}}[S_{11^{\p}}-S_{21^{\p}}-S_{13}S_{31^{\p}}+S_{31^{\p}}S_{23}]\bigg\}\left(\frac{1}{2}\frac{1}{\bx_{13}^{2}}+\frac{1}{2}\frac{1}{\bx_{1^{\p}3}^{2}}-\frac{1}{2}\frac{\bx_{11^{\p}}^{2}}{\bx_{13}^{2}\bx_{1^{\p}3}^{2}}\right).
\end{align}

%%%%%%%%%%%%%%%%%%%%
%%%%%%%%%%%%%%%%%%%%
\begin{align}
 & \frac{\dd\sigma_{3\times1}^{T}}{\dd^{2}\bp\,\dd y_{1}}=-2\frac{e^{2}g^{2}Q^{2}N_{c}}{(2\pi)^{8}}z_{1}^{2}z_{2}(z_{1}^{2}+z_{2}^{2})\int\frac{\dd z}{z}\int\dd^{8}\bx\frac{\bx_{12}\cdot\bx_{1^{\p}2}}{|\bx_{12}||\bx_{1^{\p}2}|}K_{1}(|\bx_{12}|Q_{1})K_{1}(|\bx_{1^{\p}2}|Q_{1})e^{i\bp\cdot\bx_{11^{\p}}}\nonumber \\
 & \bigg\{ C_{F}[S_{31}S_{1^{\p}3}-S_{1^{\p}3}S_{32}-S_{21}+1]-\frac{1}{2N_{c}}[S_{1^{\p}1}-S_{1^{\p}2}-S_{31}S_{1^{\p}3}+S_{1^{\p}3}S_{32}]\bigg\}\left(\frac{1}{2}\frac{1}{\bx_{13}^{2}}+\frac{1}{2}\frac{1}{\bx_{1^{\p}3}^{2}}-\frac{1}{2}\frac{\bx_{11^{\p}}^{2}}{\bx_{13}^{2}\bx_{1^{\p}3}^{2}}\right).
\end{align}

%%%%%%%%%%%%%%%%%%%%
%\newpage
%%%%%%%%%%%%%%%%%%%%
\begin{align}
 & \frac{\dd\sigma_{1\times4}^{T}}{\dd^{2}\bp\,\dd y_{1}}=2\frac{e^{2}g^{2}Q^{2}N_{c}}{(2\pi)^{8}}z_{1}^{2}z_{2}(z_{1}^{2}+z_{2}^{2})\int\frac{\dd z}{z}\int\dd^{8}\bx\frac{\bx_{12}\cdot\bx_{1^{\p}2}}{|\bx_{12}||\bx_{1^{\p}2}|}K_{1}(|\bx_{12}|Q_{1})K_{1}(|\bx_{1^{\p}2}|Q_{1})e^{i\bp\cdot\bx_{1^{\p}1}}\nonumber \\
 & \bigg\{ C_{F}[S_{13}S_{31^{\p}}-S_{31^{\p}}S_{23}-S_{12}+1]-\frac{1}{2N_{c}}[S_{11^{\p}}-S_{21^{\p}}-S_{13}S_{31^{\p}}+S_{31^{\p}}S_{23}]\bigg\}\left(\frac{1}{2}\frac{1}{\bx_{13}^{2}}+\frac{1}{2}\frac{1}{\bx_{23}^{2}}-\frac{1}{2}\frac{\bx_{12}^{2}}{\bx_{13}^{2}\bx_{23}^{2}}\right).
\end{align}

%%%%%%%%%%%%%%%%%%%%
%%%%%%%%%%%%%%%%%%%% 
\begin{align}
 & \frac{\dd\sigma_{4\times1}^{T}}{\dd^{2}\bp\,\dd y_{1}}=2\frac{e^{2}g^{2}Q^{2}N_{c}}{(2\pi)^{8}}z_{1}^{2}z_{2}(z_{1}^{2}+z_{2}^{2})\int\frac{\dd z}{z}\int\dd^{8}\bx\frac{\bx_{12}\cdot\bx_{1^{\p}2}}{|\bx_{12}||\bx_{1^{\p}2}|}K_{1}(|\bx_{12}|Q_{1})K_{1}(|\bx_{1^{\p}2}|Q_{1})e^{i\bp\cdot\bx_{11^{\p}}}\nonumber \\
 & \bigg\{ C_{F}[S_{31}S_{1^{\p}3}-S_{1^{\p}3}S_{32}-S_{21}+1]-\frac{1}{2N_{c}}[S_{1^{\p}1}-S_{1^{\p}2}-S_{31}S_{1^{\p}3}+S_{1^{\p}3}S_{32}]\bigg\}\left(\frac{1}{2}\frac{1}{\bx_{13}^{2}}+\frac{1}{2}\frac{1}{\bx_{23}^{2}}-\frac{1}{2}\frac{\bx_{12}^{2}}{\bx_{13}^{2}\bx_{23}^{2}}\right).
\end{align}

\subsubsection{Virtual corrections}

\begin{align}
 & \frac{\dd\sigma_{5}^{T}}{\dd^{2}\bp\,\dd y_{1}}=2\frac{e^{2}g^{2}Q^{2}N_{c}}{(2\pi)^{8}}z_{1}^{2}z_{2}(z_{1}^{2}+z_{2}^{2})\int_{0}^{z_{1}}\frac{\dd z}{z}\int\dd^{8}\bx\frac{\bx_{12}\cdot\bx_{1^{\p}2}}{|\bx_{12}||\bx_{1^{\p}2}|}K_{1}(|\bx_{12}|Q_{1})K_{1}(|\bx_{1^{\p}2}|Q_{1})e^{i\bp\cdot\bx_{1^{\p}1}}\nonumber \\
 & \bigg\{ C_{F}[S_{13}S_{31^{\p}}-S_{32}S_{13}-S_{21^{\p}}+1]-\frac{1}{2N_{c}}[S_{11^{\p}}-S_{12}-S_{13}S_{31^{\p}}+S_{32}S_{13}]\bigg\}\frac{1}{\bx_{31}^{2}}.
\end{align}

%%%%%%%%%%%%%%%%%%%%
%%%%%%%%%%%%%%%%%%%%
\begin{align}
 & \frac{\dd\sigma_{5}^{\star,T}}{\dd^{2}\bp\,\dd y_{1}}=2\frac{e^{2}g^{2}Q^{2}N_{c}}{(2\pi)^{8}}z_{1}^{2}z_{2}(z_{1}^{2}+z_{2}^{2})\int_{0}^{z_{1}}\frac{\dd z}{z}\int\dd^{8}\bx\frac{\bx_{12}\cdot\bx_{1^{\p}2}}{|\bx_{12}||\bx_{1^{\p}2}|}K_{1}(|\bx_{12}|Q_{1})K_{1}(|\bx_{1^{\p}2}|Q_{1})\,e^{i\bp\cdot\bx_{11^{\p}}}\nonumber \\
 & \bigg\{ C_{F}[S_{31}S_{1^{\p}3}-S_{23}S_{31}-S_{1^{\p}2}+1]-\frac{1}{2N_{c}}[S_{1^{\p}1}-S_{21}-S_{31}S_{1^{\p}3}+S_{23}S_{31}]\bigg\}\frac{1}{\bx_{31}^{2}}.
\end{align}

%%%%%%%%%%%%%%%%%%%%
%\newpage
%%%%%%%%%%%%%%%%%%%%
\begin{align}
 & \frac{\dd\sigma_{7}^{T}}{\dd^{2}\bp\,\dd y_{1}}=-2\frac{e^{2}g^{2}Q^{2}N_{c}}{(2\pi)^{8}}z_{1}^{2}z_{2}(z_{1}^{2}+z_{2}^{2})\int_{0}^{z_{1}}\frac{\dd z}{z}\int\dd^{8}\bx\,\frac{\bx_{12}\cdot\bx_{1^{\p}2}}{|\bx_{12}||\bx_{1^{\p}2}|}K_{1}(|\bx_{12}|Q_{1})K_{1}(|\bx_{1^{\p}2}|Q_{1})\,e^{i\bp\cdot\bx_{1^{\p}1}}\nonumber \\
 & \bigg\{ C_{F}[S_{13}S_{31^{\p}}-S_{32}S_{13}-S_{21^{\p}}+1]-\frac{1}{2N_{c}}[S_{11^{\p}}-S_{12}-S_{13}S_{31^{\p}}+S_{32}S_{13}]\bigg\}\left(\frac{1}{2}\frac{1}{\bx_{13}^{2}}+\frac{1}{2}\frac{1}{\bx_{23}^{2}}-\frac{1}{2}\frac{\bx_{12}^{2}}{\bx_{13}^{2}\bx_{23}^{2}}\right).
\end{align}

%%%%%%%%%%%%%%%%%%%%
%%%%%%%%%%%%%%%%%%%%
\begin{align}
 & \frac{\dd\sigma_{7}^{\star,T}}{\dd^{2}\bp\,\dd y_{1}}=-2\frac{e^{2}g^{2}Q^{2}N_{c}}{(2\pi)^{8}}z_{1}^{2}z_{2}(z_{1}^{2}+z_{2}^{2})\int_{0}^{z_{1}}\frac{\dd z}{z}\int\dd^{8}\bx\,\frac{\bx_{12}\cdot\bx_{1^{\p}2}}{|\bx_{12}||\bx_{1^{\p}2}|}K_{1}(|\bx_{12}|Q_{1})K_{1}(|\bx_{1^{\p}2}|Q_{1})\,e^{i\bp\cdot\bx_{11^{\p}}}\nonumber \\
 & \bigg\{ C_{F}[S_{31}S_{1^{\p}3}-S_{23}S_{31}-S_{1^{\p}2}+1]-\frac{1}{2N_{c}}[S_{1^{\p}1}-S_{21}-S_{31}S_{1^{\p}3}+S_{23}S_{31}]\bigg\}\left(\frac{1}{2}\frac{1}{\bx_{13}^{2}}+\frac{1}{2}\frac{1}{\bx_{23}^{2}}-\frac{1}{2}\frac{\bx_{12}^{2}}{\bx_{13}^{2}\bx_{23}^{2}}\right).
\end{align}

%%%%%%%%%%%%%%%%%%%%
%\newpage
%%%%%%%%%%%%%%%%%%%%
\begin{align}
 & \frac{\dd\sigma_{9}^{T}}{\dd^{2}\bp\,\dd y_{1}}=-\frac{e^{2}g^{2}Q^{2}N_{c}}{(2\pi)^{8}}z_{1}^{2}z_{2}(z_{1}^{2}+z_{2}^{2})\int_{0}^{z_{1}}\frac{\dd z}{z}\int\dd^{8}\bx\frac{\bx_{12}\cdot\bx_{1^{\p}2}}{|\bx_{12}||\bx_{1^{\p}2}|}K_{1}(|\bx_{12}|Q_{1})K_{1}(|\bx_{1^{\p}2}|Q_{1})e^{i\bp\cdot\bx_{1^{\p}1}}\nonumber \\
 & C_{F}\bigg[S_{11^{\p}}-S_{12}-S_{21^{\p}}+1\bigg]\frac{1}{\bx_{3}^{2}}.
\end{align}

%%%%%%%%%%%%%%%%%%%%
%%%%%%%%%%%%%%%%%%%%
\begin{align}
 & \frac{\dd\sigma_{9}^{\star,T}}{\dd^{2}\bp\,\dd y_{1}}=-\frac{e^{2}g^{2}Q^{2}N_{c}}{(2\pi)^{8}}z_{1}^{2}z_{2}(z_{1}^{2}+z_{2}^{2})\int_{0}^{z_{1}}\frac{\dd z}{z}\int\dd^{8}\bx\frac{\bx_{12}\cdot\bx_{1^{\p}2}}{|\bx_{12}||\bx_{1^{\p}2}|}K_{1}(|\bx_{12}|Q_{1})K_{1}(|\bx_{1^{\p}2}|Q_{1})e^{i\bp\cdot\bx_{11^{\p}}}\nonumber \\
 & C_{F}\bigg[S_{1^{\p}1}-S_{21}-S_{1^{\p}2}+1\bigg]\frac{1}{\bx_{3}^{2}}.
\end{align}

%%%%%%%%%%%%%%%%%%%%
%\newpage
%%%%%%%%%%%%%%%%%%%%

\begin{align}
 & \frac{\dd\sigma_{11}^{T}}{\dd^{2}\bp\,\dd y_{1}}=-2\frac{e^{2}g^{2}Q^{2}N_{c}}{(2\pi)^{8}}z_{1}^{2}z_{2}(z_{1}^{2}+z_{2}^{2})\int_{0}^{z_{1}}\frac{\dd z}{z}\int\dd^{8}\bx\frac{\bx_{12}\cdot\bx_{1^{\p}2}}{|\bx_{12}||\bx_{1^{\p}2}|}K_{1}(|\bx_{12}|Q_{1})K_{1}(|\bx_{1^{\p}2}|Q_{1})\,e^{i\bp\cdot\bx_{1^{\p}1}}\nonumber \\
 & C_{F}\bigg[S_{11^{\p}}-S_{12}-S_{21^{\p}}+1\bigg]\frac{1}{\bx_{3}^{2}}.
\end{align}

%%%%%%%%%%%%%%%%%%%%
%%%%%%%%%%%%%%%%%%%%

\begin{align}
 & \frac{\dd\sigma_{11}^{\star,T}}{\dd^{2}\bp\,\dd y_{1}}=-2\frac{e^{2}g^{2}Q^{2}N_{c}}{(2\pi)^{8}}z_{1}^{2}z_{2}(z_{1}^{2}+z_{2}^{2})\int_{0}^{z_{1}}\frac{\dd z}{z}\int\dd^{8}\bx\frac{\bx_{12}\cdot\bx_{1^{\p}2}}{|\bx_{12}||\bx_{1^{\p}2}|}K_{1}(|\bx_{12}|Q_{1})K_{1}(|\bx_{1^{\p}2}|Q_{1})\,e^{i\bp\cdot\bx_{11^{\p}}}\nonumber \\
 & C_{F}\bigg[S_{1^{\p}1}-S_{21}-S_{1^{\p}2}+1\bigg]\frac{1}{\bx_{3}^{2}}.
\end{align}

%%%%%%%%%%%%%%%%%%%%
%\newpage
%%%%%%%%%%%%%%%%%%%%
\begin{align}
 & \frac{\dd\sigma_{12}^{T}}{\dd^{2}\bp\,\dd y_{1}}=-2\frac{e^{2}g^{2}Q^{2}N_{c}}{(2\pi)^{8}}z_{1}^{2}z_{2}(z_{1}^{2}+z_{2}^{2})\int_{0}^{z_{2}}\frac{\dd z}{z}\int\dd^{8}\bx\frac{\bx_{12}\cdot\bx_{1^{\p}2}}{|\bx_{12}||\bx_{1^{\p}2}|}K_{1}(|\bx_{12}|Q_{1})K_{1}(|\bx_{1^{\p}2}|Q_{1})\,e^{i\bp\cdot\bx_{1^{\p}1}}\nonumber \\
 & C_{F}\bigg[S_{11^{\p}}-S_{12}-S_{21^{\p}}+1\bigg]\frac{1}{\bx_{3}^{2}}.
\end{align}

%%%%%%%%%%%%%%%%%%%%
%%%%%%%%%%%%%%%%%%%%
\begin{align}
 & \frac{\dd\sigma_{12}^{\star,T}}{\dd^{2}\bp\,\dd y_{1}}=-2\frac{e^{2}g^{2}Q^{2}N_{c}}{(2\pi)^{5}}z_{1}^{2}z_{2}(z_{1}^{2}+z_{2}^{2})\int_{0}^{z_{2}}\frac{\dd z}{z}\int\dd^{8}\bx\frac{\bx_{12}\cdot\bx_{1^{\p}2}}{|\bx_{12}||\bx_{1^{\p}2}|}K_{1}(|\bx_{12}|Q_{1})K_{1}(|\bx_{1^{\p}2}|Q_{1})\,e^{i\bp\cdot\bx_{11^{\p}}}\nonumber \\
 & C_{F}\bigg[S_{1^{\p}1}-S_{21}-S_{1^{\p}2}+1\bigg]\frac{1}{\bx_{3}^{2}}.
\end{align}

%%%%%%%%%%%%%%%%%%%%
%\newpage
%%%%%%%%%%%%%%%%%%%%
\begin{align}
 & \frac{\dd\sigma_{13(1)}^{T}}{\dd^{2}\bp\,\dd y_{1}}=\frac{e^{2}g^{2}Q^{2}N_{c}}{(2\pi)^{8}}z_{1}^{2}z_{2}(z_{1}^{2}+z_{2}^{2})\int_{0}^{z_{2}}\frac{\dd z}{z}\int\dd^{8}\bx\,\frac{\bx_{12}\cdot\bx_{1^{\p}2}}{|\bx_{12}||\bx_{1^{\p}2}|}K_{1}(|\bx_{12}|Q_{1})K_{1}(|\bx_{1^{\p}2}|Q_{1})\,e^{i\bp\cdot\bx_{1^{\p}1}}\nonumber \\
 & \,\bigg\{ C_{F}[S_{21^{\p}}S_{12}+1-S_{12}-S_{21^{\p}}]-\frac{1}{2N_{c}}\left(S_{11^{\p}}-S_{21^{\p}}S_{12}\right)\bigg\}\left(\frac{1}{2}\frac{1}{\bx_{13}^{2}}+\frac{1}{2}\frac{1}{\bx_{23}^{2}}-\frac{1}{2}\frac{\bx_{12}^{2}}{\bx_{13}^{2}\bx_{23}^{2}}\right).
\end{align}

%%%%%%%%%%%%%%%%%%%%
%%%%%%%%%%%%%%%%%%%%
\begin{align}
 & \frac{\dd\sigma_{13(1)}^{\star,T}}{\dd^{2}\bp\,\dd y_{1}}=\frac{e^{2}g^{2}Q^{2}N_{c}}{(2\pi)^{8}}z_{1}^{2}z_{2}(z_{1}^{2}+z_{2}^{2})\int_{0}^{z_{2}}\frac{\dd z}{z}\int\dd^{8}\bx\,\frac{\bx_{12}\cdot\bx_{1^{\p}2}}{|\bx_{12}||\bx_{1^{\p}2}|}K_{1}(|\bx_{12}|Q_{1})K_{1}(|\bx_{1^{\p}2}|Q_{1})\,e^{i\bp\cdot\bx_{11^{\p}}}\nonumber \\
 & \bigg\{ C_{F}[S_{1^{\p}2}S_{21}+1-S_{21}-S_{1^{\p}2}]-\frac{1}{2N_{c}}\left(S_{1^{\p}1}-S_{1^{\p}2}S_{21}\right)\bigg\}\left(\frac{1}{2}\frac{1}{\bx_{13}^{2}}+\frac{1}{2}\frac{1}{\bx_{23}^{2}}-\frac{1}{2}\frac{\bx_{12}^{2}}{\bx_{13}^{2}\bx_{23}^{2}}\right).
\end{align}

%%%%%%%%%%%%%%%%%%%%
%\newpage
%%%%%%%%%%%%%%%%%%%%
\begin{align}
 & \frac{\dd\sigma_{13(2)}^{T}}{\dd^{2}\bp\,\dd y_{1}}=\frac{e^{2}g^{2}Q^{2}N_{c}}{(2\pi)^{8}}z_{1}^{2}z_{2}(z_{1}^{2}+z_{2}^{2})\int_{0}^{z_{1}}\frac{\dd z}{z}\int\dd^{8}\bx\,\frac{\bx_{12}\cdot\bx_{1^{\p}2}}{|\bx_{12}||\bx_{1^{\p}2}|}K_{1}(|\bx_{12}|Q_{1})K_{1}(|\bx_{1^{\p}2}|Q_{1})\,e^{i\bp\cdot\bx_{1^{\p}1}}\nonumber \\
 & \,\bigg\{ C_{F}[S_{21^{\p}}S_{12}+1-S_{12}-S_{21^{\p}}]-\frac{1}{2N_{c}}\left(S_{11^{\p}}-S_{21^{\p}}S_{12}\right)\bigg\}\left(\frac{1}{2}\frac{1}{\bx_{13}^{2}}+\frac{1}{2}\frac{1}{\bx_{23}^{2}}-\frac{1}{2}\frac{\bx_{12}^{2}}{\bx_{13}^{2}\bx_{23}^{2}}\right).
\end{align}

%%%%%%%%%%%%%%%%%%%%
%%%%%%%%%%%%%%%%%%%%
\begin{align}
 & \frac{\dd\sigma_{13(2)}^{\star,T}}{\dd^{2}\bp\,\dd y_{1}}=\frac{e^{2}g^{2}Q^{2}N_{c}}{(2\pi)^{6}}z_{1}^{2}z_{2}(z_{1}^{2}+z_{2}^{2})\int_{0}^{z_{1}}\frac{\dd z}{z}\int\dd^{8}\bx\,\frac{\bx_{12}\cdot\bx_{1^{\p}2}}{|\bx_{12}||\bx_{1^{\p}2}|}K_{1}(|\bx_{12}|Q_{1})K_{1}(|\bx_{1^{\p}2}|Q_{1})\,e^{i\bp\cdot\bx_{11^{\p}}}\nonumber \\
 & \bigg\{ C_{F}[S_{1^{\p}2}S_{21}+1-S_{21}-S_{1^{\p}2}]-\frac{1}{2N_{c}}\left(S_{1^{\p}1}-S_{1^{\p}2}S_{21}\right)\bigg\}\left(\frac{1}{2}\frac{1}{\bx_{13}^{2}}+\frac{1}{2}\frac{1}{\bx_{23}^{2}}-\frac{1}{2}\frac{\bx_{12}^{2}}{\bx_{13}^{2}\bx_{23}^{2}}\right).
\end{align}

%%%%%%%%%%%%%%%%%%%%
%\newpage
%%%%%%%%%%%%%%%%%%%%
\begin{align}
 & \frac{\dd\sigma_{14(1)}^{T}}{\dd^{2}\bp\,\dd y_{1}}=\frac{e^{2}g^{2}Q^{2}N_{c}}{(2\pi)^{8}}z_{1}^{2}z_{2}(z_{1}^{2}+z_{2}^{2})\int_{0}^{z_{1}}\frac{\dd z}{z}\int\dd^{8}\bx\,\frac{\bx_{12}\cdot\bx_{1^{\p}2}}{|\bx_{12}||\bx_{1^{\p}2}|}K_{1}(|\bx_{12}|Q_{1})K_{1}(|\bx_{1^{\p}2}|Q_{1})\,e^{i\bp\cdot\bx_{1^{\p}1}}\nonumber \\
 & C_{F}\bigg[S_{11^{\p}}-S_{12}-S_{21^{\p}}+1\bigg]\left[\frac{1}{\bx_{31}^{2}}+\frac{1}{\bx_{32}^{2}}-\frac{1}{2}\frac{\bx_{12}^{2}}{\bx_{31}^{2}\bx_{32}^{2}}\right].
\end{align}

%%%%%%%%%%%%%%%%%%%%
%%%%%%%%%%%%%%%%%%%%
\begin{align}
 & \frac{\dd\sigma_{14(1)}^{\star,T}}{\dd^{2}\bp\,\dd y_{1}}=\frac{e^{2}g^{2}Q^{2}N_{c}}{(2\pi)^{8}}z_{1}^{2}z_{2}(z_{1}^{2}+z_{2}^{2})\int_{0}^{z_{1}}\frac{\dd z}{z}\int\dd^{8}\bx\frac{\bx_{12}\cdot\bx_{1^{\p}2}}{|\bx_{12}||\bx_{1^{\p}2}|}K_{1}(|\bx_{12}|Q_{1})K_{1}(|\bx_{1^{\p}2}|Q_{1})\,e^{i\bp\cdot\bx_{11^{\p}}}\nonumber \\
 & C_{F}\bigg[S_{1^{\p}1}-S_{21}-S_{1^{\p}2}+1\bigg]\left[\frac{1}{\bx_{31}^{2}}+\frac{1}{\bx_{32}^{2}}-\frac{1}{2}\frac{\bx_{12}^{2}}{\bx_{31}^{2}\bx_{32}^{2}}\right].
\end{align}

%%%%%%%%%%%%%%%%%%%%
%\newpage
%%%%%%%%%%%%%%%%%%%%

\begin{align}
 & \frac{\dd\sigma_{14(2)}^{T}}{\dd^{2}\bp\,\dd y_{1}}=\frac{e^{2}g^{2}Q^{2}N_{c}}{(2\pi)^{8}}z_{1}^{2}z_{2}(z_{1}^{2}+z_{2}^{2})\int_{0}^{z_{2}}\frac{\dd z}{z}\int\dd^{8}\bx\frac{\bx_{12}\cdot\bx_{1^{\p}2}}{|\bx_{12}||\bx_{1^{\p}2}|}K_{1}(|\bx_{12}|Q_{1})K_{1}(|\bx_{1^{\p}2}|Q_{1})\,e^{i\bp\cdot\bx_{1^{\p}1}}\nonumber \\
 & C_{F}\bigg[S_{11^{\p}}-S_{12}-S_{21^{\p}}+1\bigg]\left[\frac{1}{\bx_{31}^{2}}+\frac{1}{\bx_{32}^{2}}-\frac{1}{2}\frac{\bx_{12}^{2}}{\bx_{31}^{2}\bx_{32}^{2}}\right].
\end{align}

%%%%%%%%%%%%%%%%%%%%
%%%%%%%%%%%%%%%%%%%%

\begin{align}
 & \frac{\dd\sigma_{14(2)}^{\star,T}}{\dd^{2}\bp\,\dd y_{1}}=\frac{e^{2}g^{2}Q^{2}N_{c}}{(2\pi)^{8}}z_{1}^{2}z_{2}(z_{1}^{2}+z_{2}^{2})\int_{0}^{z_{2}}\frac{\dd z}{z}\int\dd^{8}\bx\frac{\bx_{12}\cdot\bx_{1^{\p}2}}{|\bx_{12}||\bx_{1^{\p}2}|}K_{1}(|\bx_{12}|Q_{1})K_{1}(|\bx_{1^{\p}2}|Q_{1})\,e^{i\bp\cdot\bx_{11^{\p}}}\nonumber \\
 & C_{F}\bigg[S_{1^{\p}1}-S_{21}-S_{1^{\p}2}+1\bigg]\left[\frac{1}{\bx_{31}^{2}}+\frac{1}{\bx_{32}^{2}}-\frac{1}{2}\frac{\bx_{12}^{2}}{\bx_{31}^{2}\bx_{32}^{2}}\right].
\end{align}

Note: In the $z \rightarrow 0$ limit there is further cancellation,
\begin{align}
\sigma_{1\times 2} +  \sigma_{13(1)} + \sigma_{13(2)} &= 0 \\
\sigma_{2\times 1} +  \sigma_{13(1)}^{*} + \sigma_{13(2)}^{*} &= 0 
\end{align}
This cancellation happens for any $\bk$ ($\bx_3$) so these terms do not contribute to BK/JIMWLK.

\bibliography{mybib_New}
\bibliographystyle{apsrev}

%\bibliography{sidis-transverse/mybib_New}

\end{document}